\documentstyle[12pt,graphicx,pslatex,ulem]{article}
\begin{document}

\begin{center}
{\Large {\bf Toroidal modeling of interaction between resistive wall mode and plasma flow
}}

Yueqiang Liu$^1$ and Youwen Sun$^2$ 
\end{center}

\vspace{-3mm}
{\it  
$^1$Euratom/CCFE Fusion Association, Culham Science Centre, Abingdon, OX14 3DB, UK\\
$^2$ Institute of Plasma Physics, Chinese Academy of Sciences, 
PO Box 1126, Hefei 230031, China
}

E-mail contact of the main author: yueqiang.liu@ccfe.ac.uk

{\small
{\bf Abstract.}
The non-linear interplay between the resistive wall mode (RWM) and the
toroidal plasma flow is numerically investigated in a full toroidal
geometry, by simultaneously solving the initial value problems for the
$n=1$ RWM and the $n=0$ toroidal force balance equation. Here $n$ is
the toroidal mode number. The neoclassical toroidal viscous torque is
identified as the major momentum sink that brakes the toroidal
plasma flow during the non-linear evolution of the RWM. This holds for
a mode that is initially either unstable or stable. For an initially
stable RWM, the braking of the flow, and hence the eventual growth of
the mode, depends critically on the initial perturbation amplitude.    
}

\section{Introduction}
The resistive wall mode (RWM) is a global, macroscopic
magnetohydrodynamic (MHD) instability that often limits the
operational space of fusion devices \cite{ChuPPCF10}. The origin of
the mode is often an equilibrium current or pressure driven
external ideal kink instability.  It is of
particular importance to study the RWM physics and control in the so
called advanced tokamaks, which aim at achieving economic fusion
reactors. In advanced tokamaks, suppression of the RWM often leads to a
significant gain in the plasma pressure under steady state conditions,
and thus a significant gain in the eventual fusion power production.

Even though the RWM is often regarded, and in fact studied, both in
theory and experiments \cite{ChuPPCF10}, as a linear MHD mode, there is abundant
experimental evidence pointing to the non-linear interactions between
this instability and other MHD modes \cite{MatsunagaIAEA10}, energetic particles
\cite{MatsunagaNF10,OkabayashiPP11}, as well as the plasma flow
\cite{ReimerdesNF09}. This work investigates 
the last question, namely the non-linear coupling between the RWM and
the toroidal rotation of a tokamak plasma.

This question has previously been addressed in analytic theory, based
on cylindrical approximations
\cite{FitzpatrickPP02,GimblettPP04,FitzpatrickPP07}. For instance, in
Ref. \cite{FitzpatrickPP02}, a simple set of non-linear evolution equations
is derived and investigated for the RWM in a large aspect ratio,
rotating, viscous tokamak plasma. This model essentially couples the
Fitzpatrick-Aydemir dispersion relation \cite{FitzpatrickNF96} for an
RWM to a toroidal momentum balance equation. Simulations in
Ref. \cite{FitzpatrickPP02} show that a sufficiently large amplitude
of the RWM triggers the plasma deceleration and the loss of the wall
stabilization for the mode. A somewhat different RWM model
(visco-resistive Finn model \cite{FinnPP95}) was adopted in
Ref. \cite{GimblettPP04}, which essentially investigated the
non-linear interaction between an unstable resistive wall tearing mode (RWTM)
and the plasma flow, with an additional presence of an external error
field. 

In this work, we carry out toroidal simulations of the non-linear
coupling between the RWM and the plasma flow, using the recently
developed MARS-Q code \cite{Liu12a}. Our RWM model follows a single fluid,
full resistive MHD formulation in a generic toroidal geometry, with
additional damping of the mode from a strong parallel sound wave
damping model. We solve the MHD equations together with a toroidal
momentum balance equation, that includes both the electromagnetic
torque and the neoclassical toroidal viscous torque, induced by the 3D
field perturbations due to an RWM. For simplicity, we neglect the
additional effect from error fields, although the latter can be easily
included into the MARS-Q formulation. 

The toroidal formulation is introduced in Section 2. Section 3 reports
numerical results, where the non-linear interaction between an either
unstable or stable RWM, and the plasma flow, is investigated. Section
4 summarizes the work.             
 
\section{Toroidal formulation of non-linear coupling between RWM and
  plasma flow} \label{sec:form}
The formulation is essentially the same as that of the MARS-Q code
\cite{Liu12a}, which was developed for investigating the rotational
braking and the penetration of resonant magnetic perturbation
(RMP) fields into a stable plasma. Below we give a brief description
of the equations that we solve for studying the non-linear interaction between the
RWM and the plasma flow, in a generic toroidal geometry. The RWM is described
by single fluid perturbed MHD equations that incorporates a (sheared)
toroidal flow, with the angular velocity $\Omega(r)$ that depends on
the plasma minor radius $r$  
\begin{eqnarray}
(\frac{\partial}{\partial t}+in\Omega){\bf \xi}&=&{\bf v} + ({\bf\xi}\cdot\nabla\Omega)R^2\nabla\phi, \label{eq:xi}\\
\rho(\frac{\partial}{\partial t}+in\Omega){\bf v}&=&-\nabla p + {\bf j}\times{\bf B} +{\bf
  J}\times{\bf b} - \rho\left[2\Omega\nabla Z\times{\bf
  v}+({\bf
	 v}\cdot\nabla\Omega)R^2\nabla\phi\right]-\nabla\cdot\Pi, \label{eq:v}\\ 
(\frac{\partial}{\partial t}+in\Omega){\bf b}&=&\nabla\times({\bf v}\times{\bf B})+({\bf
  b}\cdot\nabla\Omega)R^2\nabla\phi - \nabla\times(\eta{\bf j}), \label{eq:b}\\
(\frac{\partial}{\partial t}+in\Omega)p&=&-{\bf v}\cdot\nabla P-\Gamma
P\nabla\cdot{\bf v}, \label{eq:p}
\end{eqnarray}
where $(R,Z,\phi)$ is the cylindrical coordinate system for the
torus. The quantities $\rho, {\bf B}, {\bf J}=\nabla\times{\bf B}, P$
denote the 
equilibrium plasma density, magnetic field, the plasma current density,
and the plasma pressure, respectively. The quantities ${\bf \xi}, {\bf v}, {\bf
  b}, {\bf j}=\nabla\times{\bf b}, p$, which are also our solution
variables, represent the plasma displacement, the perturbed velocity, magnetic field, current, 
and pressure, respectively. $n$ is the toroidal mode number. Note
that, in this work, we consider only a single $n$ ($n=1$) RWM. This
justifies the choice of the perturbed MHD equations. 
In the single fluid MHD approximation, the ratio of specific heats,
$\Gamma$ from Eq. (\ref{eq:p}), is taken to be 5/3 as for an ideal
gas. The above system of equations is written and solved in a
dimensionless form, where the time is normalized by the Alfv\'en time
at the magnetic axis, the length is normalized by the major radius
$R_0$ of the torus, the magnetic field is normalized by the toroidal vacuum
field $B_0$ at the magnetic axis, and the pressure is normalized by
$B_0^2/\mu_0$, where $\mu_0=4\pi\times10^{-7}$H/m is the vacuum
permeability.  

It is worthwhile to comment on two specific components from the above
model for the RWM. A damping term, in the form of a viscous tensor
$\Pi$, is included into the momentum equation (\ref{eq:v}). For a
proper description of the damping physics of the RWM, it is essential
to include this type of extra damping terms. In principle, ideal MHD
can provide continuum damping on the mode, both from Alfv\'en and
sound waves. The former requires the existence of rational surfaces
inside the plasma, whilst the latter occurs only in a finite pressure
plasma. However, ideal MHD theory generally does not seem to
adequately predict the mode damping observed in experiments, both on
the critical rotation velocity for the mode suppression \cite{ReimerdesPPCF07},
and on the resonant field amplification from the response of a stable
RWM \cite{LanctotPP11}. So far probably the most adequate damping model for
the RWM is the drift kinetic model, in which the viscous
tensor term 
is calculated from the drift kinetic theory, and represents the    
perturbed kinetic pressure tensor \cite{LiuPP08}. On the other hand, it
has been noted that, in many cases, the RWM damping can be modeled by
a (strong) parallel viscous force \cite{ChuPP95,LiuNF05} 
\begin{eqnarray}
\nabla\cdot\Pi=\rho\kappa_{\|}|k_{\|}v_{th,i}| 
  \left[{\bf v}+({\bf\xi}\cdot\nabla)R^2\Omega\nabla\phi\right]_{\|}, \label{eq:sdamp}
\end{eqnarray}
where $\kappa$ is a numerical
coefficient determining the damping ``strength''. $k_{\|}=(n-m/q)/R$
is the parallel wave number, with $m$ being the poloidal harmonic
number and $q$ being the safety factor. $v_{th,i}=\sqrt{2T_i/M_i}$ is
the thermal ion velocity, with $T_i, M_i$ being the thermal ion
temperature and mass, respectively. The parallel component of the perturbed velocity is taken
along the equilibrium field line. This damping model, originally
coming from a fluid closure of the ion Landau damping of the parallel
sound wave, is often used in the RWM modeling \cite{ChuPP95,LiuNF05} and in many
cases provides adequate damping on the mode.  This damping model is
used in the present work.

As for the second comment, we use a resistive plasma model to describe the
RWM in this study, as evident from Eq. (\ref{eq:b}), where $\eta$ is
the plasma resistivity. With the normalization convention adopted in our
formulation, $\eta$ is the inverse of the (magnetic) Lundquist
number. The reason for choosing a resistive plasma, is to improve the
numerical resolution for the electromagnetic torque, which generally
occurs near rational surfaces. A resistive model, with large
Lundquist number at least in the plasma core, is also more relevant
to realistic experiments. We point out though, that the electromagnetic
torque can remain finite even in an ideal plasma \cite{TaylorPRL03,LiuPP12b}.

Normally the RWM is well described by ideal 
MHD (with extra damping physics as mentioned above). Inclusion of the plasma
resistivity can have two effects on the RWM: it generally increases the
mode growth rate; and more qualitatively, it couples the RWM to a
tearing mode \cite{FinnPP95,BettiPP98}. The question of which mode becomes more
dominant depends on the ratio between the plasma resistive time and
the wall resistive time, among other physical parameters. In this
work, we assume that the wall resistivity is much larger than the
plasma resistivity (the plasma is close to an ideal plasma in the
core), so that the mode remains a predominantly RWM.       

For the RWM modeling, the above MHD equations inside the plasma are
solved together with the vacuum equations outside the plasma, as well
as a resistive wall equation with the thin shell approximation. 

In addition, a toroidal momentum balance equation
\begin{eqnarray}
\frac{\partial L}{\partial t}  = D(L) + T_{j\times b} + T_{\rm NTV}
+ T_{\rm source},  \label{eq:mom}
\end{eqnarray}
is also solved for $L=\rho<R^2>\Omega$, where $<\cdot>$ denotes the
surface average of a quantity. $D(L)$ is a (linear) momentum diffusion
operator. $T_{j\times b}$ is the surface averaged, toroidal
electromagnetic ${\bf j}\times{\bf b}$ torque, computed as 
\begin{eqnarray*}
T_{j\times b}=\oint R{\bf j}\times{\bf b}\cdot\hat\phi dS/\oint dS,
\end{eqnarray*}
where $R$ is the major radius, ${\bf j}$ and ${\bf b}$ are the (total)
perturbed plasma current and magnetic field, respectively, as the
solution of Eqs. (\ref{eq:v})-(\ref{eq:p}). 
$S$ denotes the flux surface.

 $T_{\rm NTV}$ is the neoclassical toroidal viscous
torque, computed using formulas from Ref. \cite{ShaingNF10},
where various regimes are smoothly connected. The module of the
perturbed magnetic field (the Lagrangian part), from
Eqs. (\ref{eq:v})-(\ref{eq:p}), directly enters into the NTV
computations.  $T_{\rm source}$ is the momentum source term. The RWM 
perturbations enter into the momentum sinks ($T_{j\times b}$ and
$T_{\rm NTV}$) in a quadratic form. The NTV torque, which is a
consequence of the radial drift of banana orbits of trapped thermal
particles, due to the presence of 3D magnetic fields, is generally a
rather non-linear function of the plasma flow velocity. We refer to
\cite{Liu12a} for detailed description of these sink and source terms
in the MARS-Q formulation.     

Note that the parallel viscous force, Eq. (\ref{eq:sdamp}) which
describes the damping of the $n\neq 0$ RWM, does not explicitly enter
into the $n=0$ momentum balance equation (\ref{eq:mom}).
 
An important point is that we assume a momentum balance is reached at
the start of our simulation $t=0$, such that 
$D(L(t=0)) + T_{source} = 0$. This condition essentially defines the initial
rotation frequency.  
By further assuming that the momentum source does not change during
the non-linear simulation, we instead solve the momentum balance
equation for the change of the toroidal momentum, $\Delta L=L(t)-L(0)$,
\begin{eqnarray}
\frac{\partial \Delta L}{\partial t}  = D(\Delta L)  + T_{j\times b} +
T_{NTV}, \label{eq:momf}
\end{eqnarray}
in which the momentum source does not explicitly enter into the equation. 

We use the homogeneous Neumann (Dirichlet) boundary condition for $\Delta L$ at
the plasma center (edge). The Dirichlet edge boundary condition is motivated by
discussions from Ref. \cite{Fitzpatrick93}. 

In MARS-Q, the $n=1$ single fluid MHD equations, the vacuum and wall
equations, as well as the $n=0$ toroidal momentum
balance equation are solved together as an initial value problem. An
adaptive, semi-implicit time stepping scheme is employed
\cite{Liu12a}. In particular, a fully implicit scheme is used for solving the
MHD equations, allowing a time step larger than the Alfv\'en time
without introducing numerical instability. The initial condition for
the $n=1$ perturbation is chosen as the 
eigenfunction of the linear RWM, computed by running the eigenvalue
code MARS-F \cite{Liu00}.   

\section{Numerical results}
\subsection{Equilibrium}
\begin{figure}
\begin{center}
\includegraphics[width=8cm]{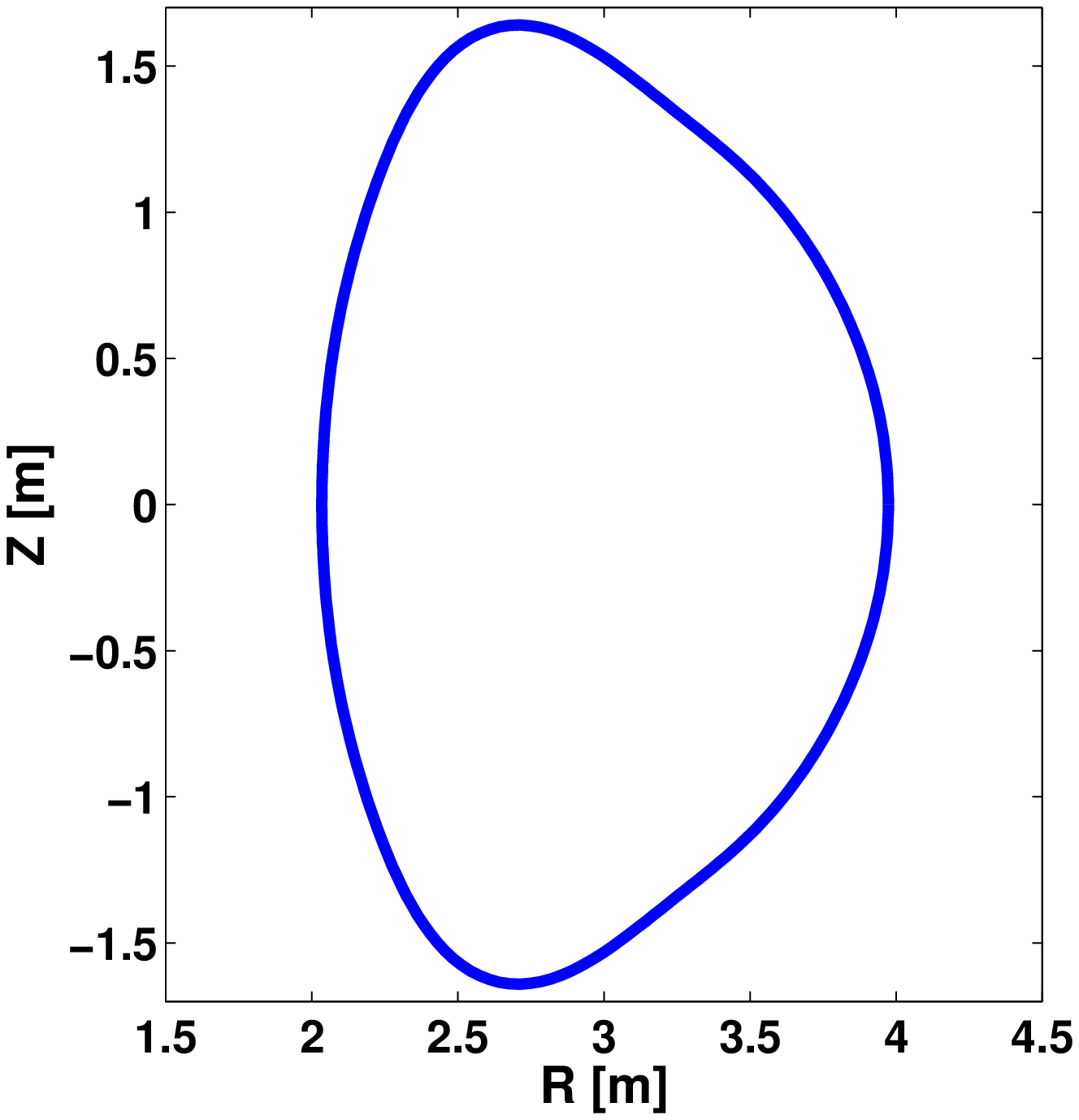}
\caption{The plasma boundary shape of the toroidal equilibrium.}  
\label{fig:bnd}
\end{center}
\end{figure}

We consider a toroidal plasma representing the so called advanced
tokamak, with the plasma boundary shape, shown in Fig. \ref{fig:bnd},
resembling a typical JET plasma (in terms of aspect ratio, elongation
and triangularity). The plasma boundary is up-down symmetric. The vacuum
toroidal field, at the magnetic axis $R_0=2.90$m, is assumed to be
$B_0=1.22$T. The total plasma current is 1.56MA.  

\begin{figure}
\begin{center}
\includegraphics[width=6.5cm]{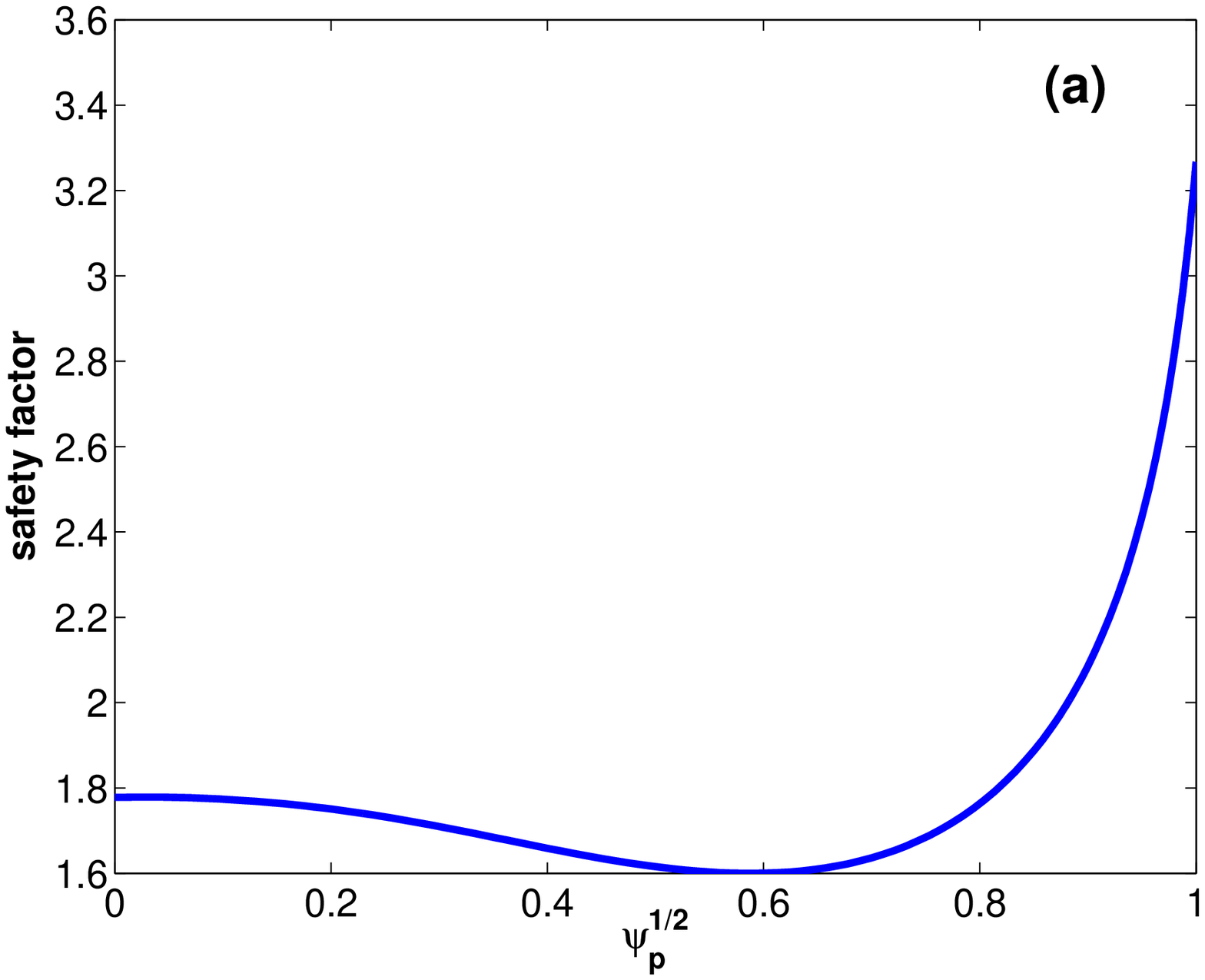}
\includegraphics[width=6.5cm]{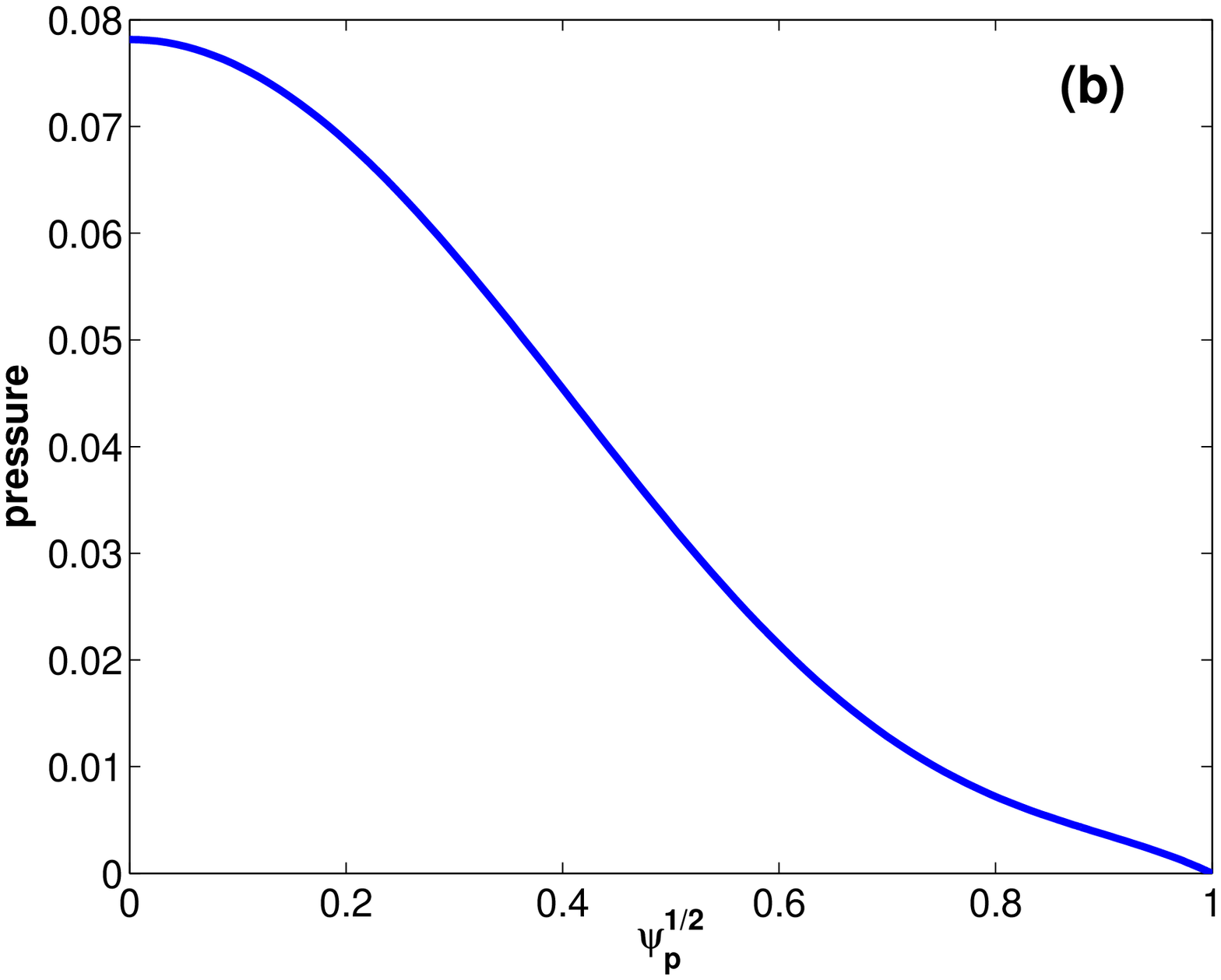}
\includegraphics[width=6.5cm]{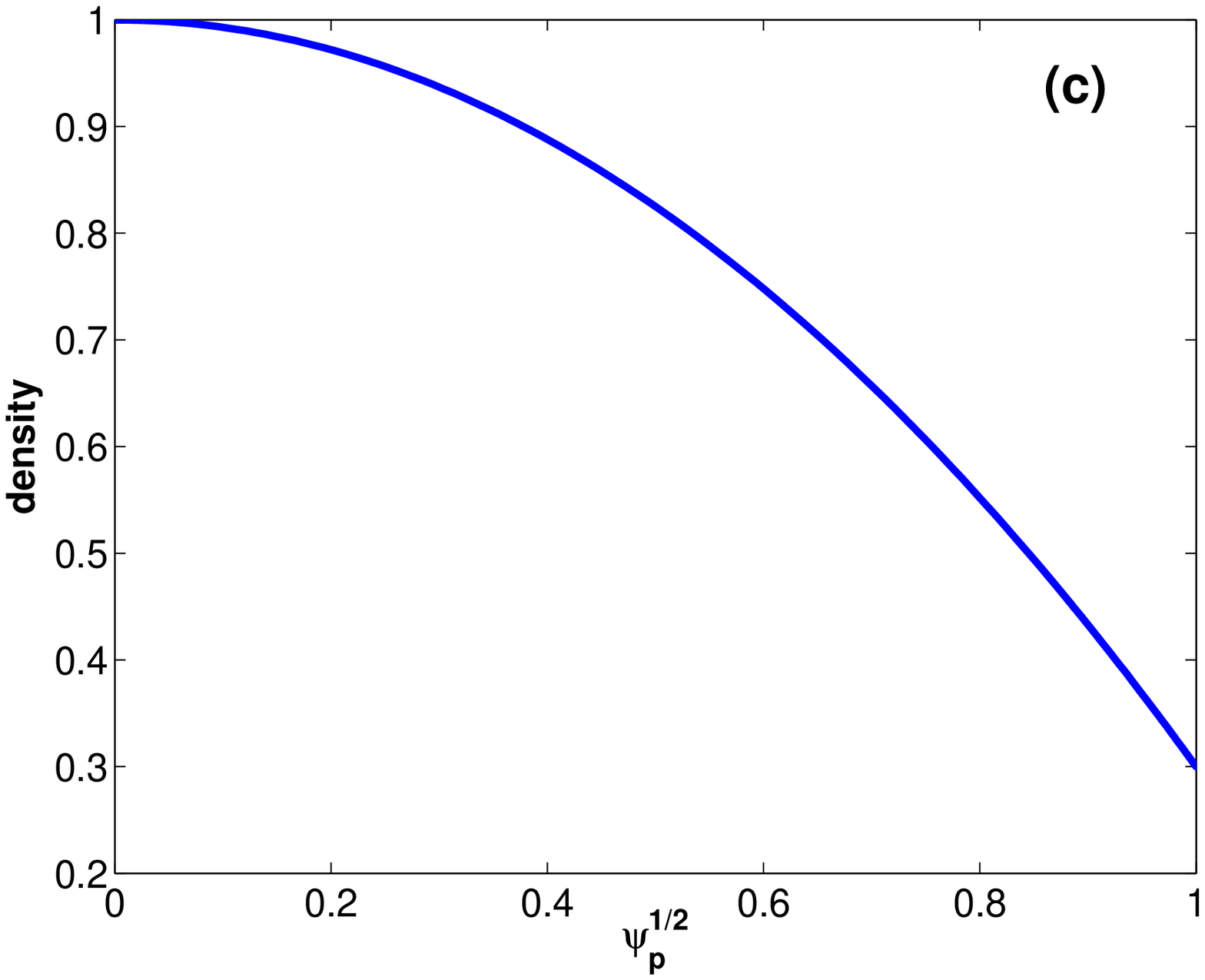}
\includegraphics[width=6.5cm]{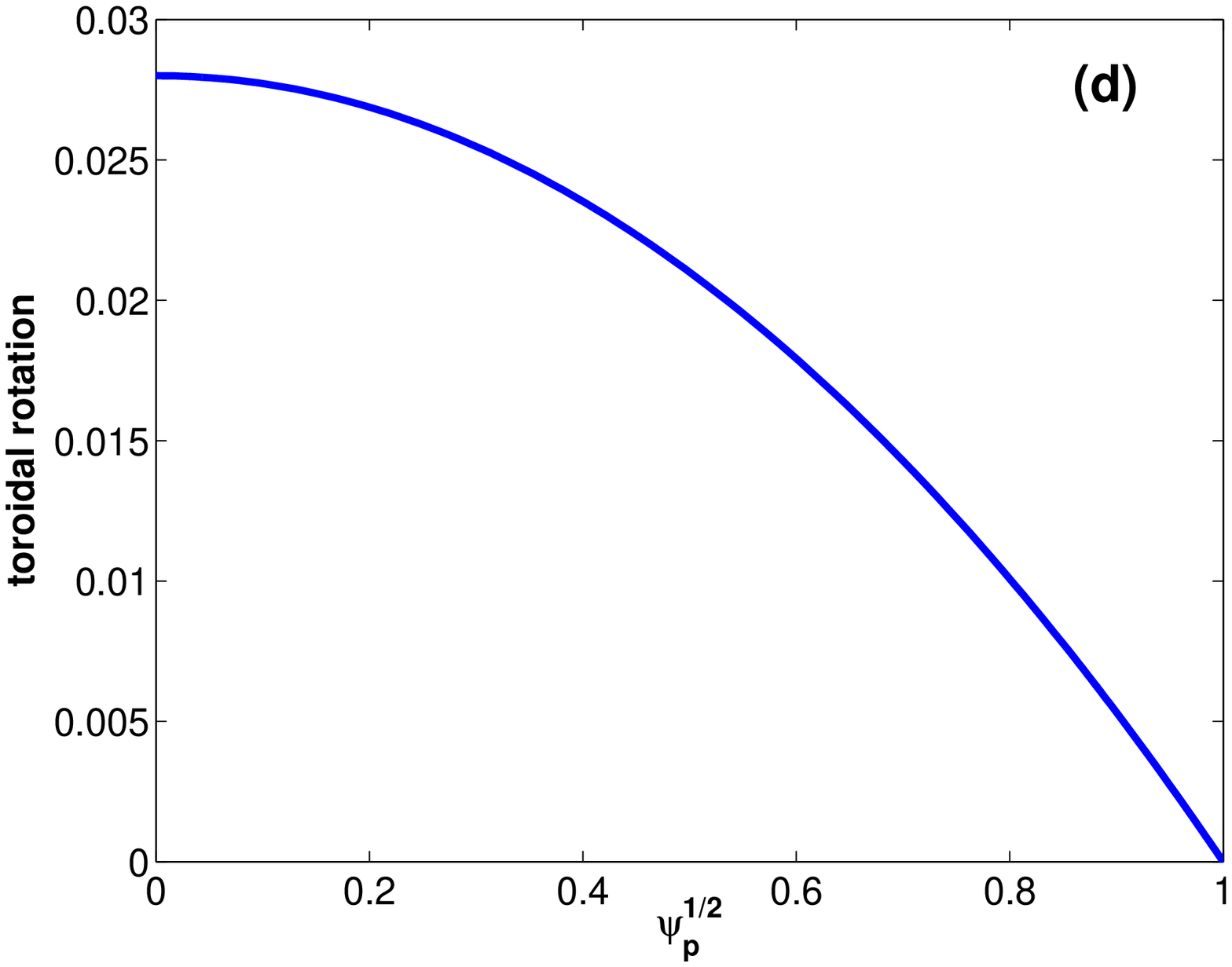}
\caption{Radial profiles of the equilibrium quantities for (a) the
  safety factor $q$, (b) the plasma pressure normalized by
  $B_0^2/\mu_0$, (c) the plasma density normalized to unity at the
  magnetic axis, and (d) the toroidal rotation frequency of the
  plasma, normalized by the on-axis Alfv\'en frequency.}  
\label{fig:eq}
\end{center}
\end{figure}

The radial profiles of the equilibrium quantities are shown in
Fig. \ref{fig:eq}.  The advanced tokamak aims at high 
beta, high fraction of non-inductive current drive. A peaked pressure
profile, as shown in Fig. \ref{fig:eq}(b), results in a large fraction
of the bootstrap current in the middle region of the plasma column,
which in turn creates a rather flat current profile and a slightly
reversed $q$ profile in the plasma core. The equilibrium with a broad
current profile and a peaked pressure profile tends to be more
susceptible to the external kink instability \cite{BondesonNF99}, which becomes an RWM in
the presence of a close fitting resistive wall. This motivates our
choice of the equilibrium for studying the interaction between the RWM
and the plasma flow. We note that this equilibrium pressure does not
have a clear edge pedestal. For the pressure driven RWM, the edge
transport barrier is not a critical factor. 

The density profile,
shown in Fig. \ref{fig:eq}(c), is normalized to unity at the magnetic
axis. The amplitude of the density only enters into defining the
Alfv\'en time in the dimensionless, single fluid MHD
equations. However, the density amplitude does enter into the NTV
calculation (via the collisionality coefficients). In this work, we
assume a thermal electron number density of
$N_{e0}=3.09\times10^{19}$m$^{-3}$. The thermal (Deuterium) ion density is the same as
that of the electrons. This yields the Alfv\'en time of $\tau_A\equiv
R_0\sqrt{\mu_0\rho_0}/B_0 = 0.86\mu$s at the magnetic axis. Here
$\rho_0\equiv N_{e0}m_i$, and $m_i$ is the mass of Deuterium ions.

\subsection{Linear stability}
\begin{figure}
\begin{center}
\includegraphics[width=6.5cm]{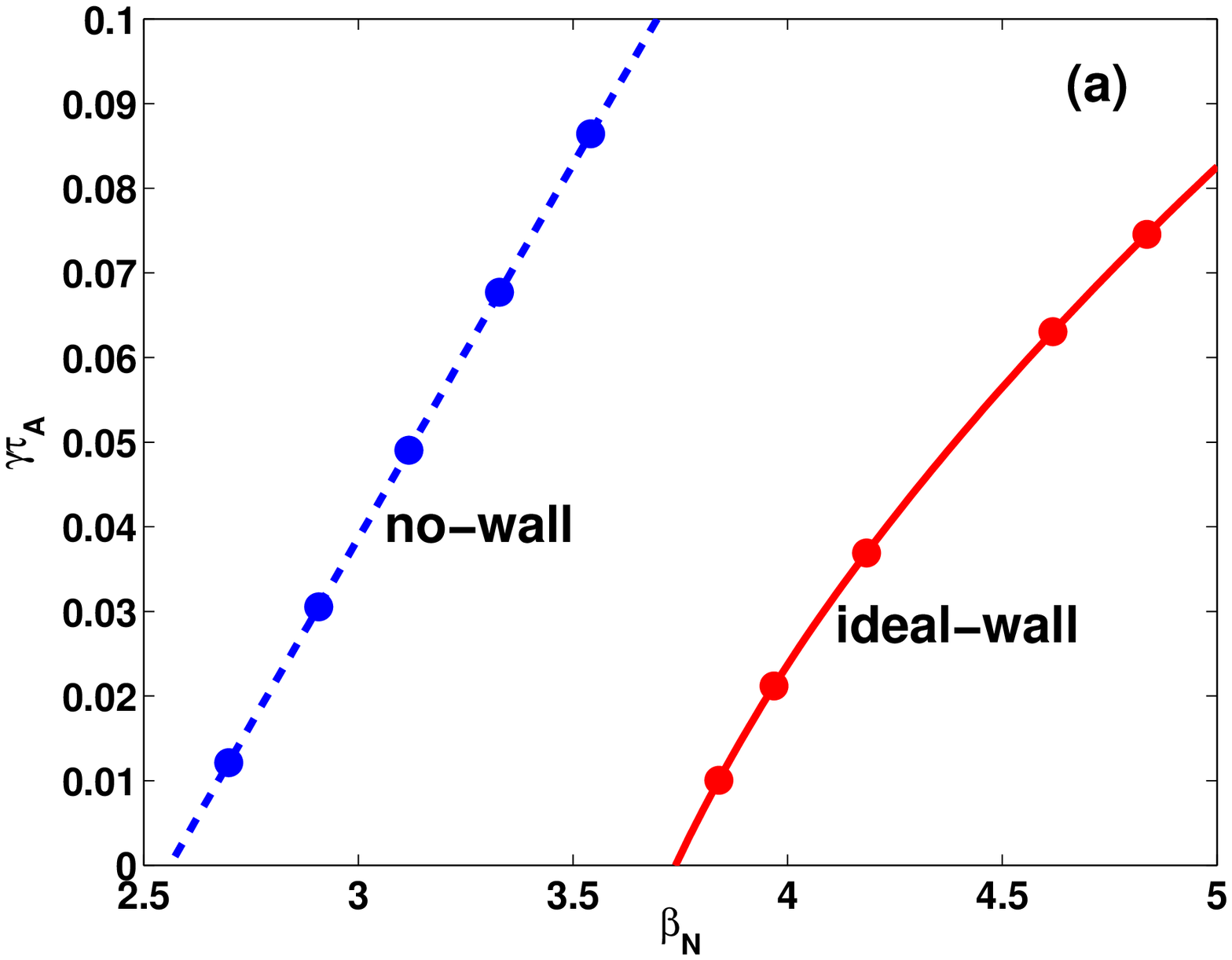}
\includegraphics[width=6.5cm]{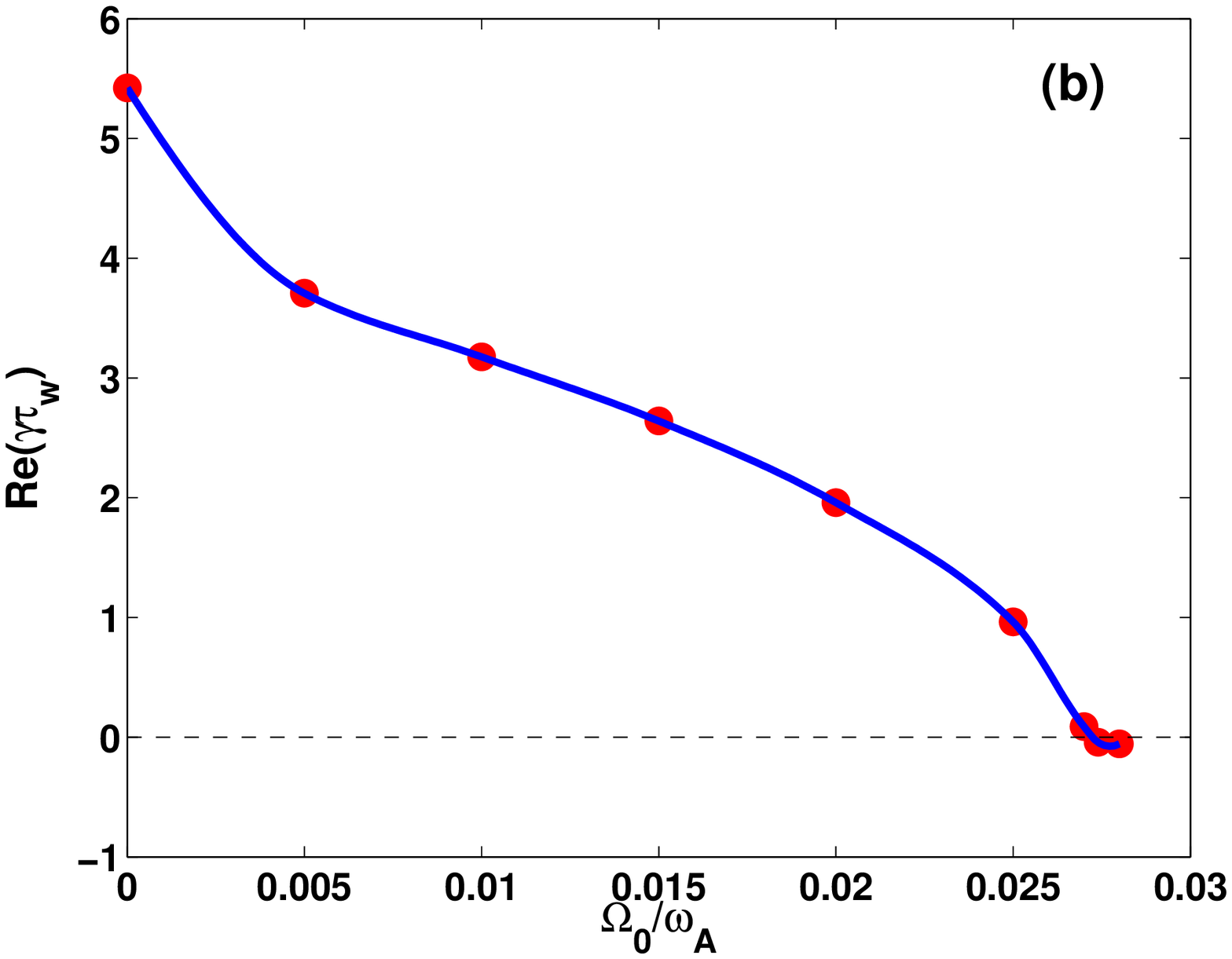}
\caption{Linear stability of (a) the ideal external, pressure-driven kink
  mode and (b) the resistive wall mode in the presence of the plasma flow.}  
\label{fig:stab}
\end{center}
\end{figure}

For the chosen equilibrium profiles, we first investigate the linear
stability of the external kink mode, using the MARS-F code. Figure
\ref{fig:stab}(a) shows the computed no-wall (dashed line) and ideal-wall
(solid line) growth rates
of the mode, while scanning the normalized plasma
pressure $\beta_N$. Only the amplitude of the plasma pressure is varied
during the scan, without changing the radial profile shown in
Fig. \ref{fig:eq}(b). The total plasma current is also fixed. An ideal, conformal
wall is located at the 1.25$a$ minor radius. The computed no-wall and
ideal-wall beta limits, which correspond to the marginal stability
points in Fig. \ref{fig:stab}(a), are $\beta_N^{\rm nw}=2.56$ and
$\beta_N^{\rm iw}=3.74$, respectively.   

Replacing the ideal wall by a resistive wall, located at the same
minor radius $r_w=1.25a$, we obtain unstable RWMs in the plasma
pressure range between the no-wall and the ideal-wall beta limits. In
this study, we choose a representative RWM at $\beta_N=3.15$, which is
half way between the no-wall and ideal-wall limits. The safety
factor is $q_0=1.78$ at the magnetic axis, $q_a=3.27$ at the
plasma edge, and $q_{95}=2.73$. Assuming a resistive plasma, with the
Lundquist number $S=10^7$ and a uniform plasma resistivity (a more
realistic resistivity profile, e.g. $S\propto T_e^{3/2}$, tends to
increase the TM contribution towards the plasma edge, which is what we
try to avoid in this work), and 
utilizing the parallel sound wave damping model (with
$\kappa_{||}=1.5$) as described in
Section \ref{sec:form}, we compute the stability of the RWM in the
presence of toroidal plasma flow. We scan the amplitude of the flow
speed while fixing the radial profile of the toroidal rotation as shown
in Fig. \ref{fig:eq}(d). The growth rate of the RWM versus the
on-axis rotation frequency $\Omega_0$ is reported in
Fig. \ref{fig:stab}(b). Note that we normalize the mode growth
rate by the wall time $\tau_w$ here. Although a fixed wall time of
$\tau_w=10^4\tau_A=8.6$ms is assumed in these computations, we mention that
the growth rate, normalized by the wall time, depends neither on the
wall time nor the Alfv\'en time, for a typical inertia-free RWM as
studied in this case. The RWM is fully stabilized by the plasma
flow, when the on-axis rotation frequency exceeds a critical value of
$\Omega_0^{\rm cr}=2.73\times10^{-2}\omega_A$, where
$\omega_A=B_0/(R_0\sqrt{\mu_0\rho_0})$ is the on-axis toroidal Alfv\'en
frequency. The damping of the mode in this case comes from the
Alfv\'en and sound wave continuum damping, as well as from the extra
ion Landau damping represented by the parallel viscous force term. 
As long as the plasma flow speed remains sub-sonic, which is the case
considered here, more resonant damping is achieved with increasing the
rotation speed, explaining the monotonic decrease of the mode's growth
rate as shown in Fig. \ref{fig:stab}(b).
 
\subsection{Interaction between initially unstable RWM and flow}
The stability analysis in Fig. \ref{fig:stab}(b) allows us to choose
two representative cases for investigating the interaction between the
RWM and the plasma flow. For the first case, reported in this
Subsection, we consider an initial
plasma rotation frequency of $\Omega_0=2.5\times 10^{-2}\omega_A$. In
this case, the RWM is still linearly unstable. We launch the MARS-Q
simulation starting from the computed linear eigenfunction at
$\Omega_0=2.5\times 10^{-2}\omega_A$. The amplitude of the
eigenfunction is scaled down to a small value, such that the perturbed
radial field amplitude is $|b^1|=3\times10^{-5}$ at the $q=2$ rational
surface and at the outboard mid-plane. 

Here $b^1$ is defined as $b^1=J{\bf B}\cdot\nabla s$, with
$J=(\nabla s\cdot\nabla\chi\times\nabla\phi)^{-1}$ being the jacobian
of the curve-linear flux coordinate system
$(s,\chi,\phi)$. $s=\sqrt{\psi_p}$ is the square root of the
normalized equilibrium poloidal flux $\psi_p$ ($\psi_p=0$ corresponds
to the magnetic axis, and $\psi_p=1$ corresponds to the plasma
edge). $\chi$ is the (generalized) poloidal angle. The perturbed
magnetic field quantities are normalized by the on axis vacuum toroidal field
$B_0$. The $b^1$ field reported in this study is always taken at the
outboard mid-plane.  

\begin{figure}
\begin{center}
\includegraphics[width=6.5cm]{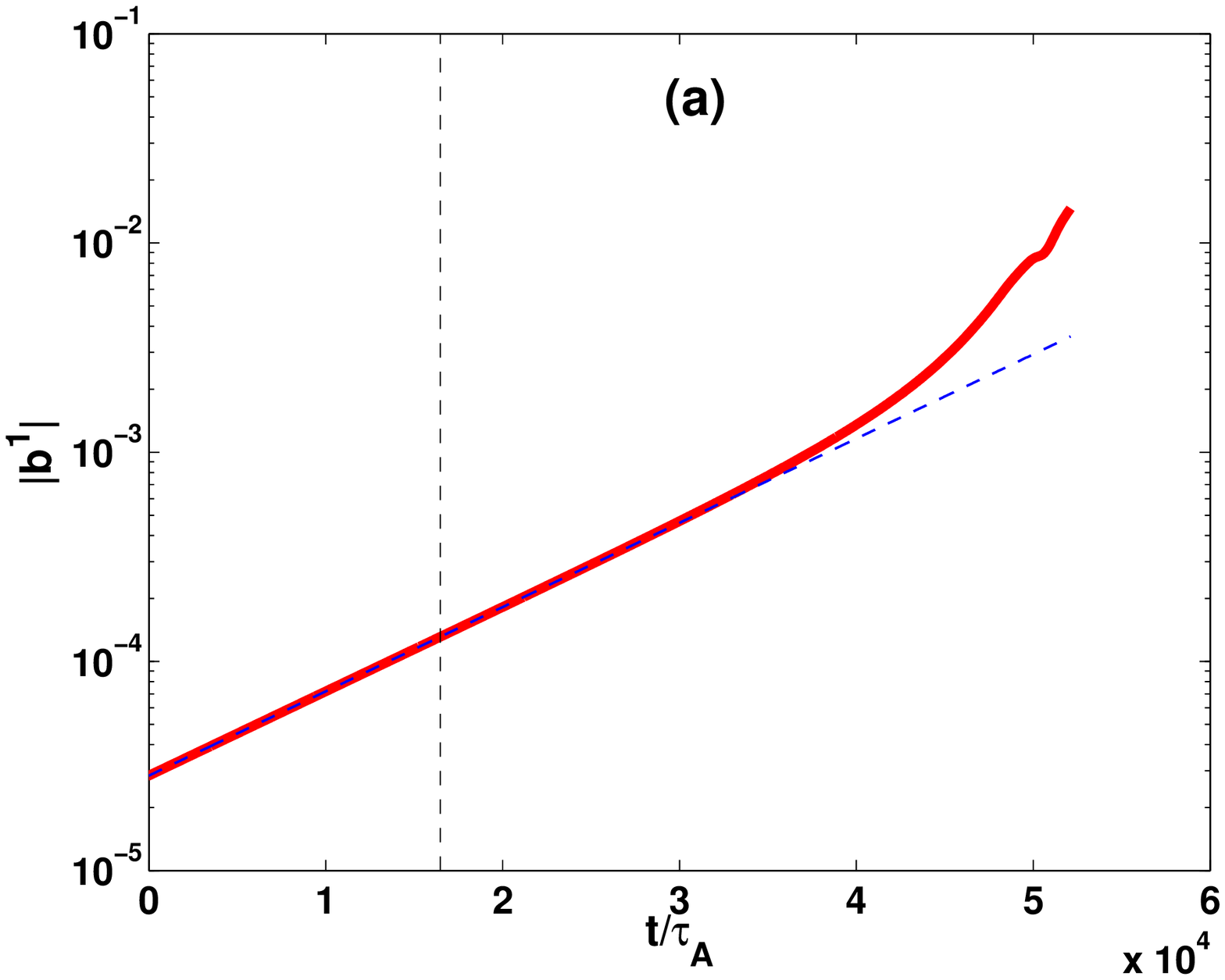}
\includegraphics[width=6.5cm]{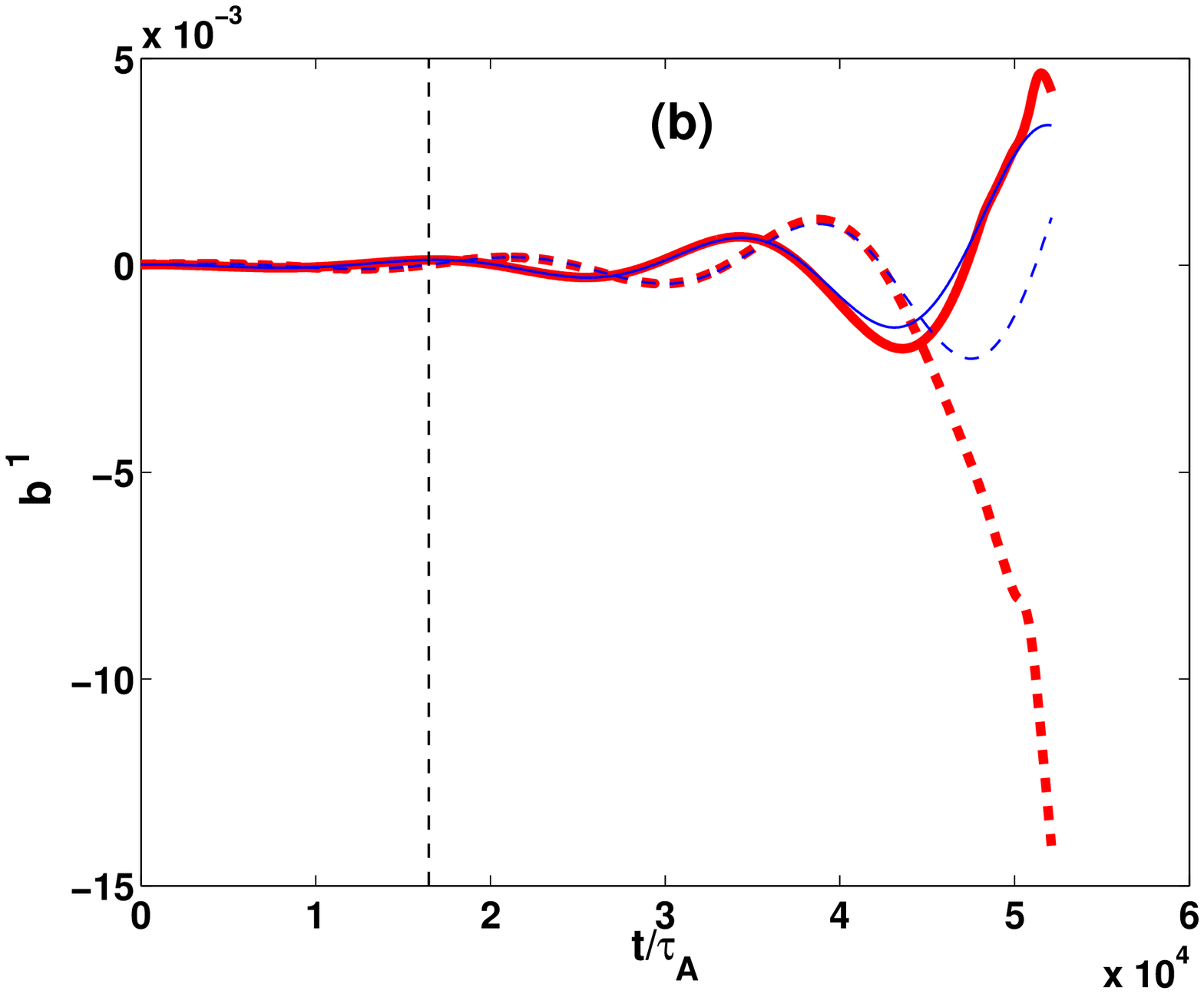}
\includegraphics[width=6.5cm]{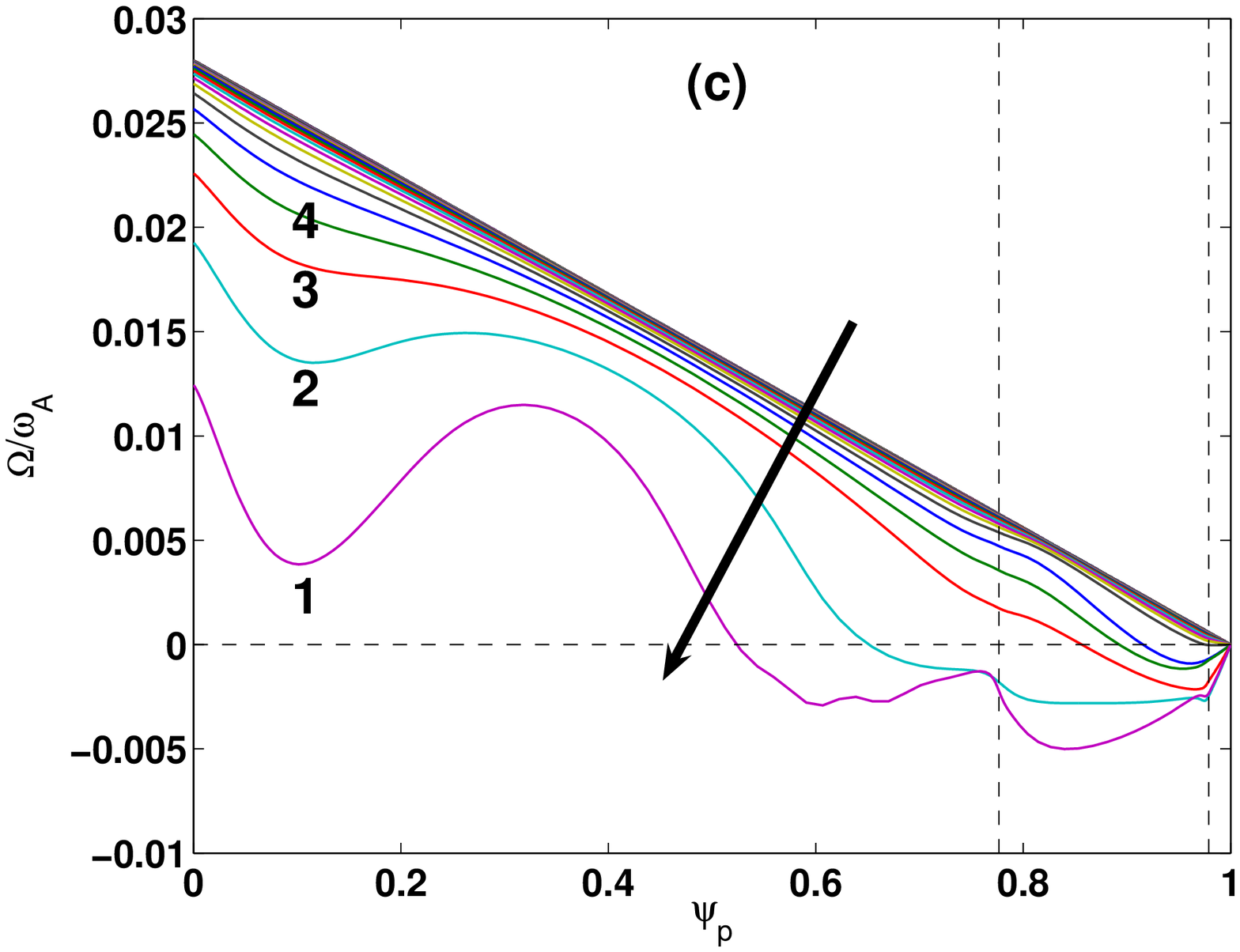}
\includegraphics[width=6.5cm]{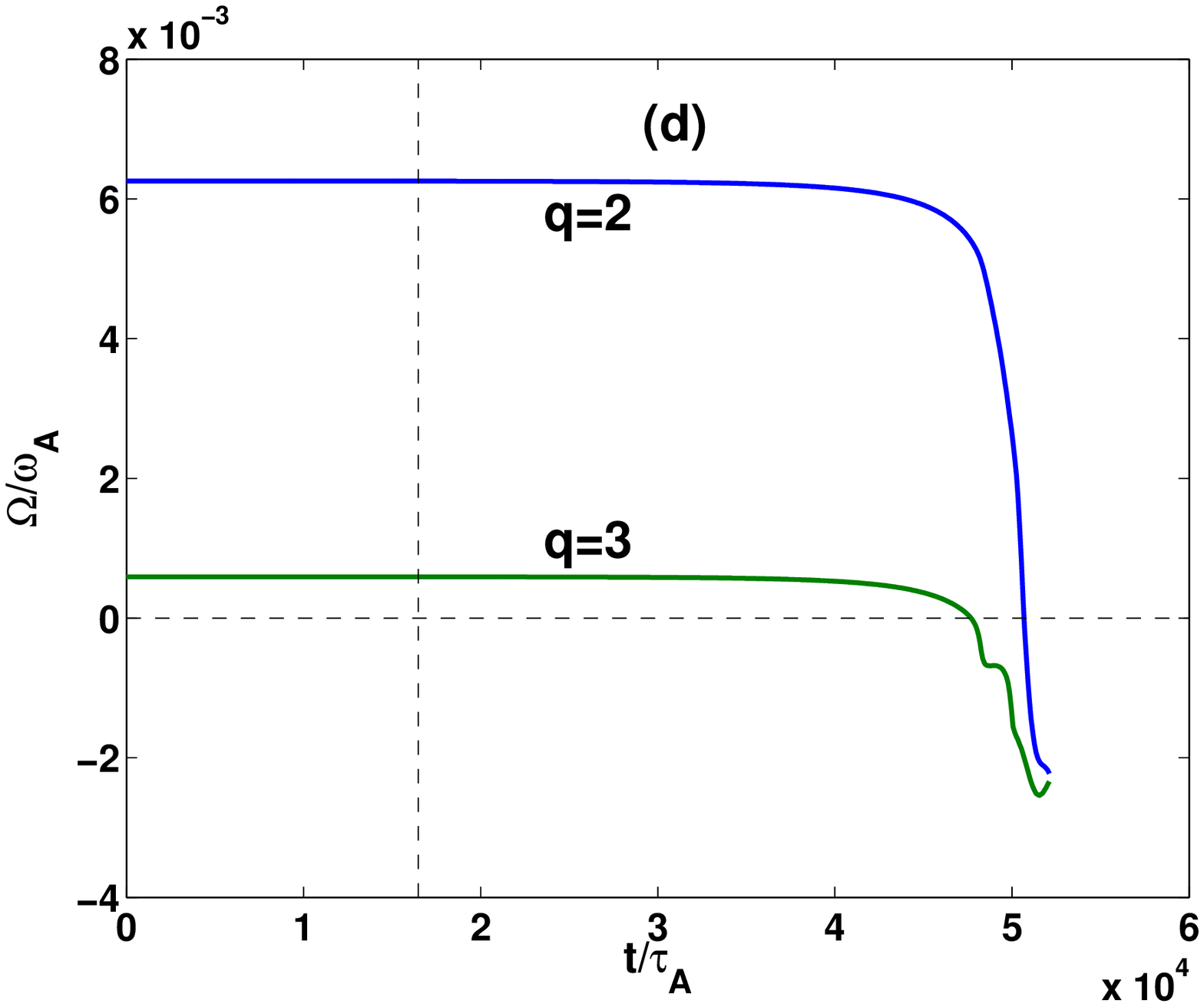}
\caption{Evolution of an initially unstable RWM: (a) the amplitude
of the perturbed radial field $b^1$ at 
the $q=2$ surface, (b) the real and imaginary parts of the perturbed
radial field $b^1$ at the $q=2$ surface, (c) the radial profile of the
plasma rotation frequency, and (d) the plasma rotation frequency at
the $q=2$ and $q=3$ surfaces. The dashed vertical lines in (a,b,d)
indicate the moment of time when the non-linear coupling between the
mode and the plasma flow is switched on. The dashed vertical lines in
(c) indicate the location of the $q=2$ and 3 rational surfaces,
respectively. The numbered lines in (c) correspond to time: 1 -
5.21$\time10^4\tau_A$, 2 - 5.13$\times10^4\tau_A$, 3 -
5.03$\times10^4\tau_A$, and 4 - 4.95$\times10^4\tau_A$. Both the electromagnetic
and the NTV torques are 
included in this simulation.}  
\label{fig:qlu}
\end{center}
\end{figure}

Figure \ref{fig:qlu} summarizes the computational results. In order to
verify whether the linear phase of the initial value code (MARS-Q)
recovers the results of the eigenvalue solver (MARS-F), we decouple
the time evolution of the RWM from the momentum solver (i.e. the
initial toroidal rotation is not changed) for the first
280 time steps, which ends at time $T_1=1.65\times10^4\tau_A$=14ms,
$\tau_A=1/\omega_A$, indicated by the vertical dashed lines in
Figs. \ref{fig:qlu}(a,b,d). The solid curve in Fig. \ref{fig:qlu}(a)
shows the amplitude of the radial field $b^1$ at the $q=2$ surface,
computed by MARS-Q. Indeed an exponential growth is recovered during
$t\in[0,T_1]$. An analytic fitting of the numerical data yields
$\gamma\tau_A=9.27\times10^{-5} + 3.58\times10^{-4}i$ for the linear
phase, agreeing well with the eigenvalue of
$\gamma\tau_A=9.63\times10^{-5} + 3.58\times10^{-4}i$ from the
eigenvalue solver MARS-F. Note that, because of the plasma flow, the
mode is rotating while growing. Figure \ref{fig:qlu}(b) shows both the
real (thick solid curve) and imaginary (thick dashed curve) parts of
$b^1(q=2)$. The analytic fitting, shown as 
dashed curve in Fig. \ref{fig:qlu}(a) and thin-line curves in
Fig. \ref{fig:qlu}(b), is
performed for these complex quantity data.      

At time $t=T_1$, we turn on the non-linear coupling between the RWM
and the plasma flow, and continue the MARS-Q simulation. In the
momentum balance equation, both the electromagnetic and the NTV
torques are included.  During this phase, the mode
becomes more unstable than the linear instability, accompanied by a
flow damping shown in Figs. \ref{fig:qlu}(c,d). Figure \ref{fig:qlu}(c) shows
the evolution of the radial profile of the toroidal rotation frequency
during the non-linear phase. In this and later simulations, we plot
the radial profiles only at about 60 time slices, equally spaced in
time. The total number of time steps for this simulation is 855
(at $t=5.22\times10^4\tau_A$),
before a full braking of the rotation occurs. Figure \ref{fig:qlu}(d)
plots the simulated time traces of the rotation frequency at the $q=2$
and $q=3$ rational surfaces. We notice that the rotation also switches
sign near the end of the simulation. This is mainly due to the
momentum sink induced by the NTV torque, as will be shown in
further analysis. The time interval between $T_1$ and the full
braking of the edge rotation (beyond the $q=2$ surface) of the plasma
is about $3.4\times10^4\tau_A=29$ms. The time period for a
considerable change of the rotation, as well as the toroidal torques
shown in Fig. \ref{fig:qlutorq} below, is about 10ms for the given
plasma parameters studied here. For a comparison, the full braking
time constant of the toroidal plasma flow is reported to be about 20ms
for a high-beta DIII-D plasma, due to the onset of the RWM
\cite{ReimerdesNF09}. The rotation braking time, observed in typical
MAST plasmas using the resonant
magnetic perturbation fields, was also reported to
be tens of milliseconds \cite{LiuPPCF12}.

The rotation braking shown in Figs. \ref{fig:qlu}(c,d) is associated
with the momentum sink terms - the electromagnetic and the NTV torques
in our model. Figure \ref{fig:qlutorq} compares time traces of the
radially integrated, net ${\bf j}\times{\bf b}$ and NTV torques acting
on the plasma column, during
the non-linear phase of the MARS-Q simulation, for the same case shown
in Fig. \ref{fig:qlu}. These net torques have negative values,
indicating deceleration of the plasma flow. More interestingly, we find
that, by amplitude, the net NTV torque is much larger than the net
electromagnetic torque. In other words, the NTV torque contributes the
major part of rotational damping in this case. In fact this holds for
all the cases considered in this paper. However, we cannot conclude that
the NTV torque is {\it always} dominant over the ${\bf j}\times{\bf
  b}$ for damping the plasma flow in a toroidal plasma.  One counter
example was reported in Ref. \cite{LiuPPCF12}, where a static RMP
(instead of a nearly static MHD mode) field was applied to a MAST
plasma, and where we observed a larger electromagnetic torque than the
NTV torque for braking the plasma flow. There are at least two factors
that affect the comparison between these two sink terms. One is the
number of rational surfaces inside the plasma. In the case reported in
Ref. \cite{LiuPPCF12}, the applied field has a high-$n$ ($n=3,4,6$)
toroidal mode number and high $q$-value, resulting in a very large
number of rational surfaces. This facilitates generation of large net
electromagnetic torque. The second factor is the initial plasma flow
speed. At fast plasma flow (as that in low aspect ratio tokamaks such
as MAST), the NTV torque tends to be small. The NTV torque becomes
considerably larger at slow plasma flow, due to a resonant
enhancement effect as discussed in \cite{LiuPPCF12} and
references therein.     

We also note that the amplitudes of both the net electromagnetic
and the NTV torques, shown in Fig. \ref{fig:qlutorq}, grow faster
than the mode amplitude. In fact, during the initial phase of the
non-linear interaction, say between $t_1=T_1$ and
$t_2=3\times10^4\tau_A$ when the mode still grows nearly linearly,
the amplitudes of the net torques also grow nearly linearly, with
the numerically recovered growth rates of about
$\gamma_{jxb}\tau_A=1.83\times10^{-4}$ and
$\gamma_{NTV}\tau_A=1.88\times10^{-4}$ for the net ${\bf j}\times{\bf
  b}$ and NTV torques, respectively. These growth rates are about
twice larger than that of the linear RWM, confirming the quadratic nature of
the torques.
 
The amplitudes of the torques shown in Fig. \ref{fig:qlutorq} can be
converted into the SI unit by multiplying the dimensionless results by
a factor $R_0^3B_0^2/\mu_0=1.3\times10^7$, yielding total torques of
up to tens of Nm for this case.  We point out that, since both the
electromagnetic and the NTV torques are quadratic functions of the
mode amplitude, for an unstable RWM, the torque can become large as
the mode amplitude grows to a large value. For a stable RWM at smaller
amplitude, both torques can be order of magnitude smaller as shown in
the next Subsection.  Several Nm's torque is measured in JET plasmas  
by applying an $n=1$ RMP fields \cite{Sun10}. Dedicate experiments
have also been carried out 
in DIII-D \cite{Garofalo11}, where the measured NTV torque is in the
order of several Nm. The direct modeling of this DIII-D plasma using MARS-Q
also recovers well the experimental value \cite{LiuPPCF12}.

\begin{figure}
\begin{center}
\includegraphics[width=10cm]{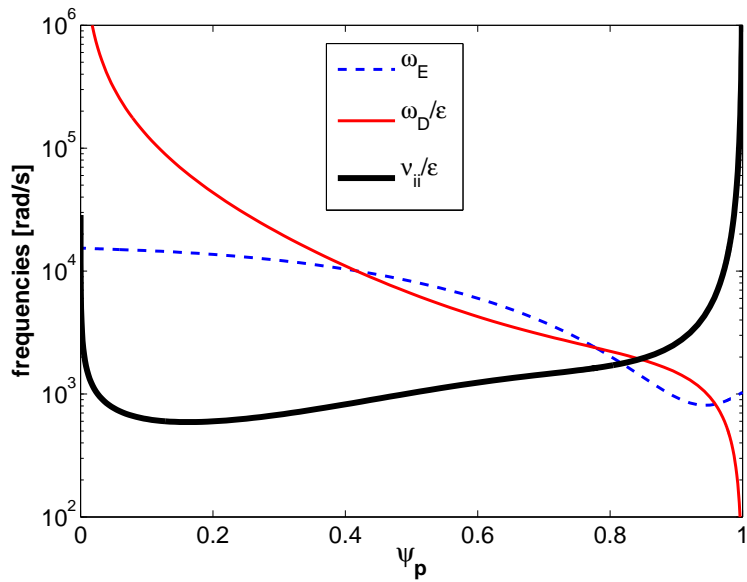}
\caption{Radial profiles of the initial (i.e. before closing the
	 non-linear coupling loop) ${\bf E}\times{\bf B}$
	 rotation frequency $\omega_E$, the precessional drift frequency
	 $\omega_D$ of deeply
	 trapped thermal ions at thermal velocity, and the ion-ion
	 collision frequency $\nu_{ii}$. $\epsilon$ is approximately the
	 inverse aspect ratio.}  
\label{fig:freqcmp}
\end{center}
\end{figure}

Figure \ref{fig:freqcmp} shows important frequencies relevant to
the NTV torque calculation, for the plasma studied in this work. In
the core region, both the ion-ion collision frequency and 
the ${\bf E}\times{\bf B}$ frequency (shown at the initial stage)
are small compared to the precessional drift frequency,
resulting in a predominantly resonant NTV torque in the so-called
superbanana or superbanana plateau regime. This resonant torque can be
large, and is responsible for the fast decay of rotation in the
plasma core, as shown in Fig. \ref{fig:qlu}(c).  
In the middle of the plasma column, the ${\bf E}\times{\bf B}$
rotation frequency is larger than the precession and collision frequencies of
trapped ions. The resulting NTV torque has a predominant non-resonant
component (in the so-called $\nu-\sqrt{\nu}$ regime), which is
normally small. In the region close to the plasma edge, where $\omega_E$ is
comparable or smaller than $\omega_D/\epsilon$, we expect again a resonant
NTV torque which is larger. Very close
to the plasma edge, the collision frequency dominates, resulting in
the so-called $1/\nu$-regime, where the NTV torque is again relatively
small.

\begin{figure}
\begin{center}
\includegraphics[width=6.5cm]{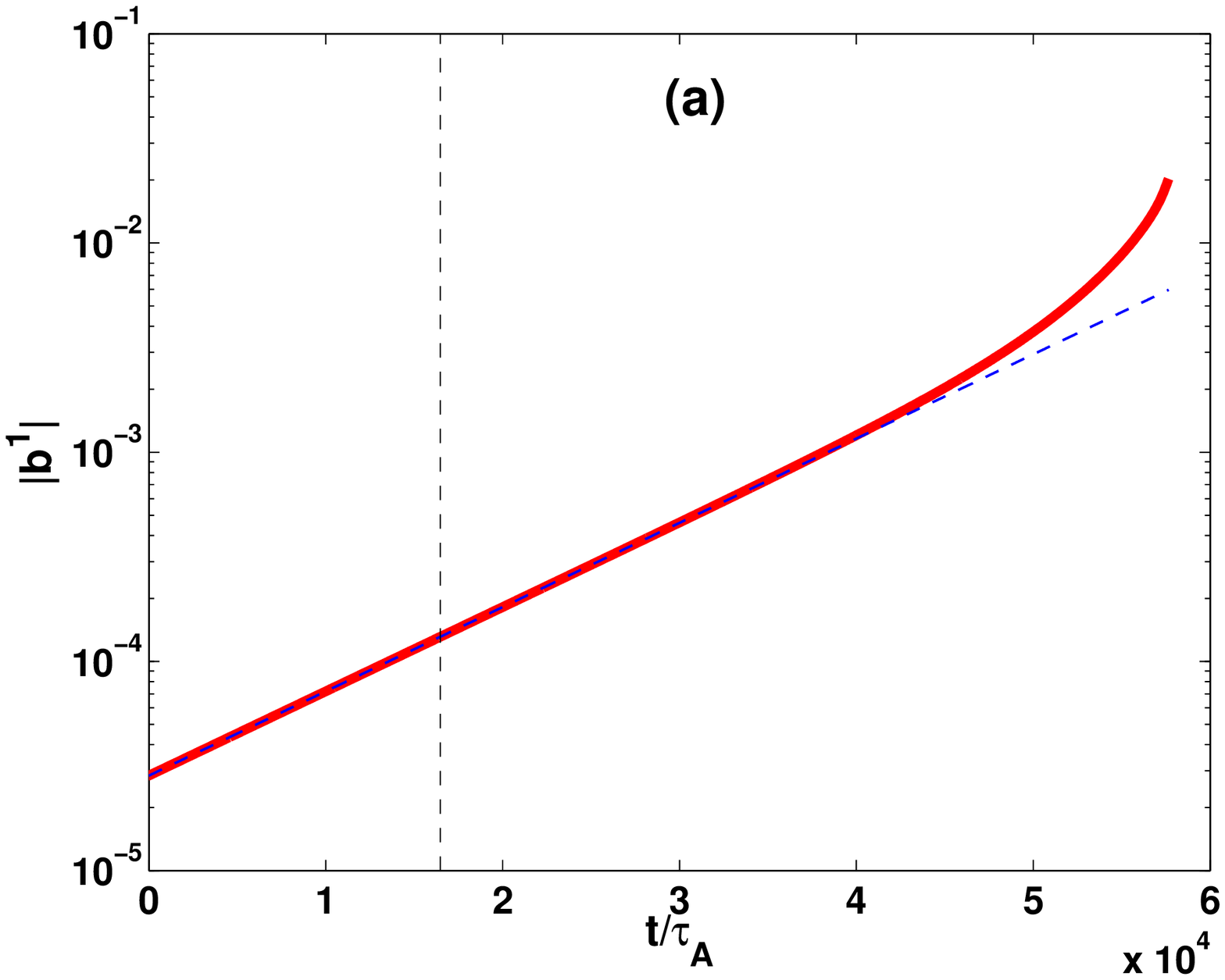}
\includegraphics[width=6.5cm]{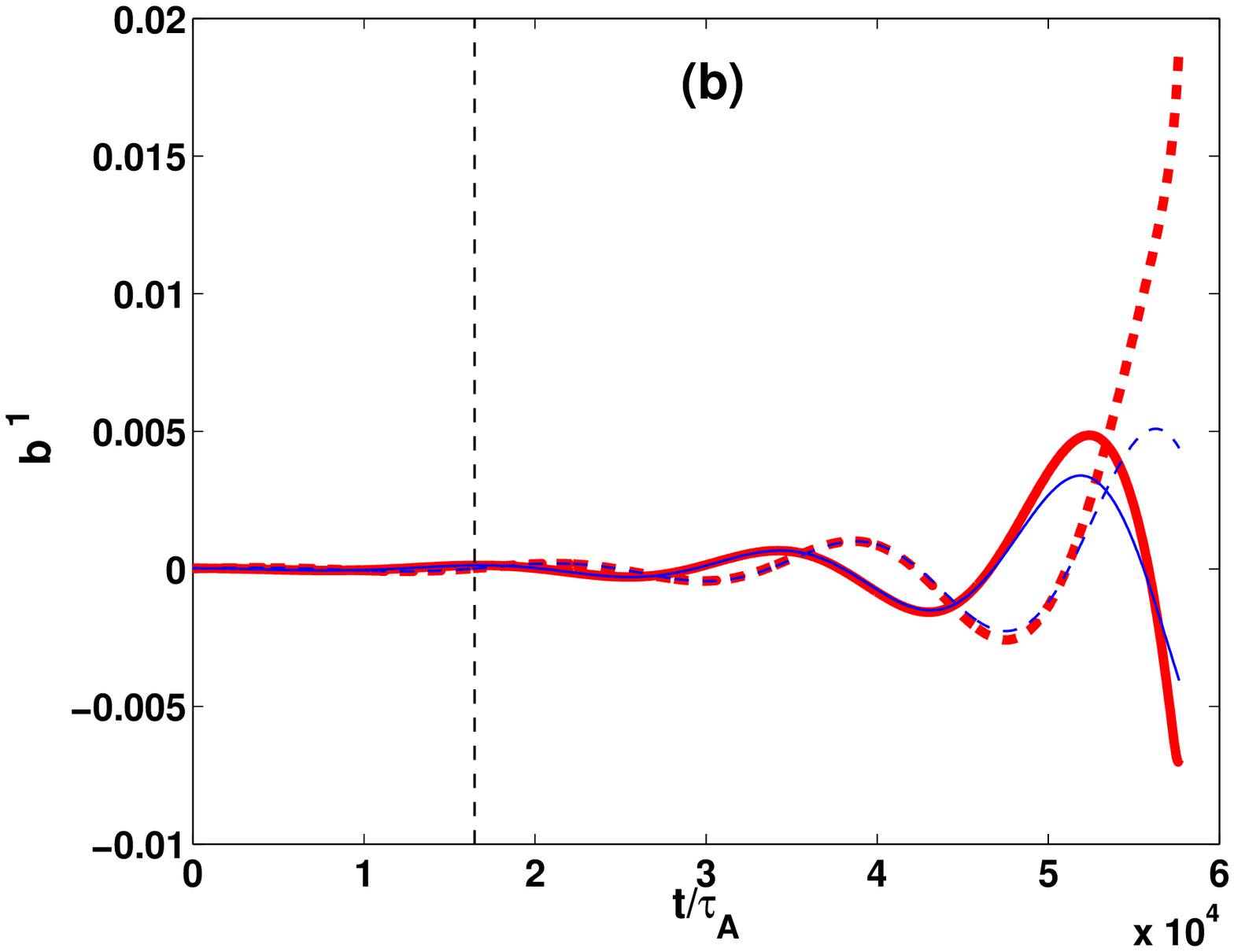}
\includegraphics[width=6.5cm]{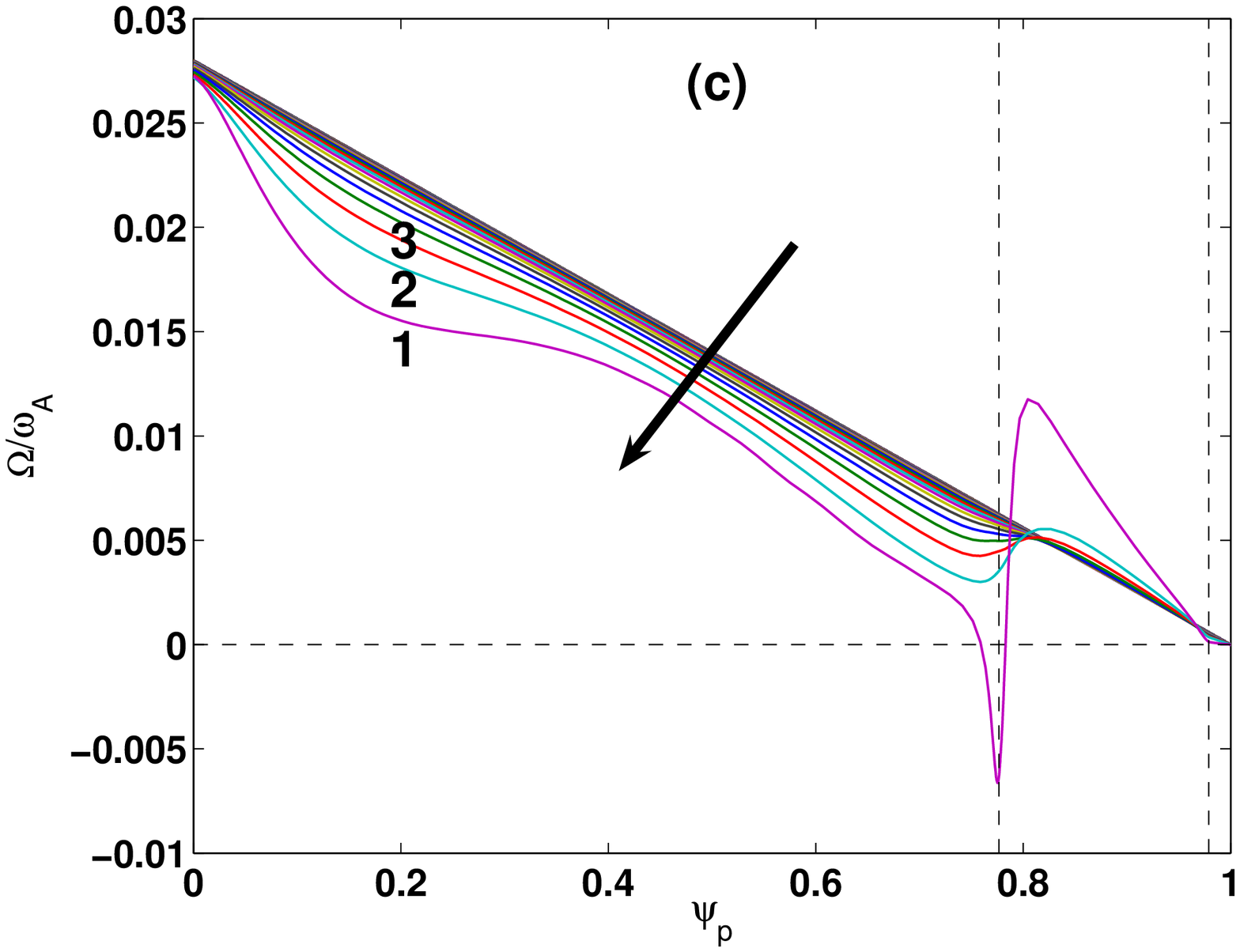}
\includegraphics[width=6.5cm]{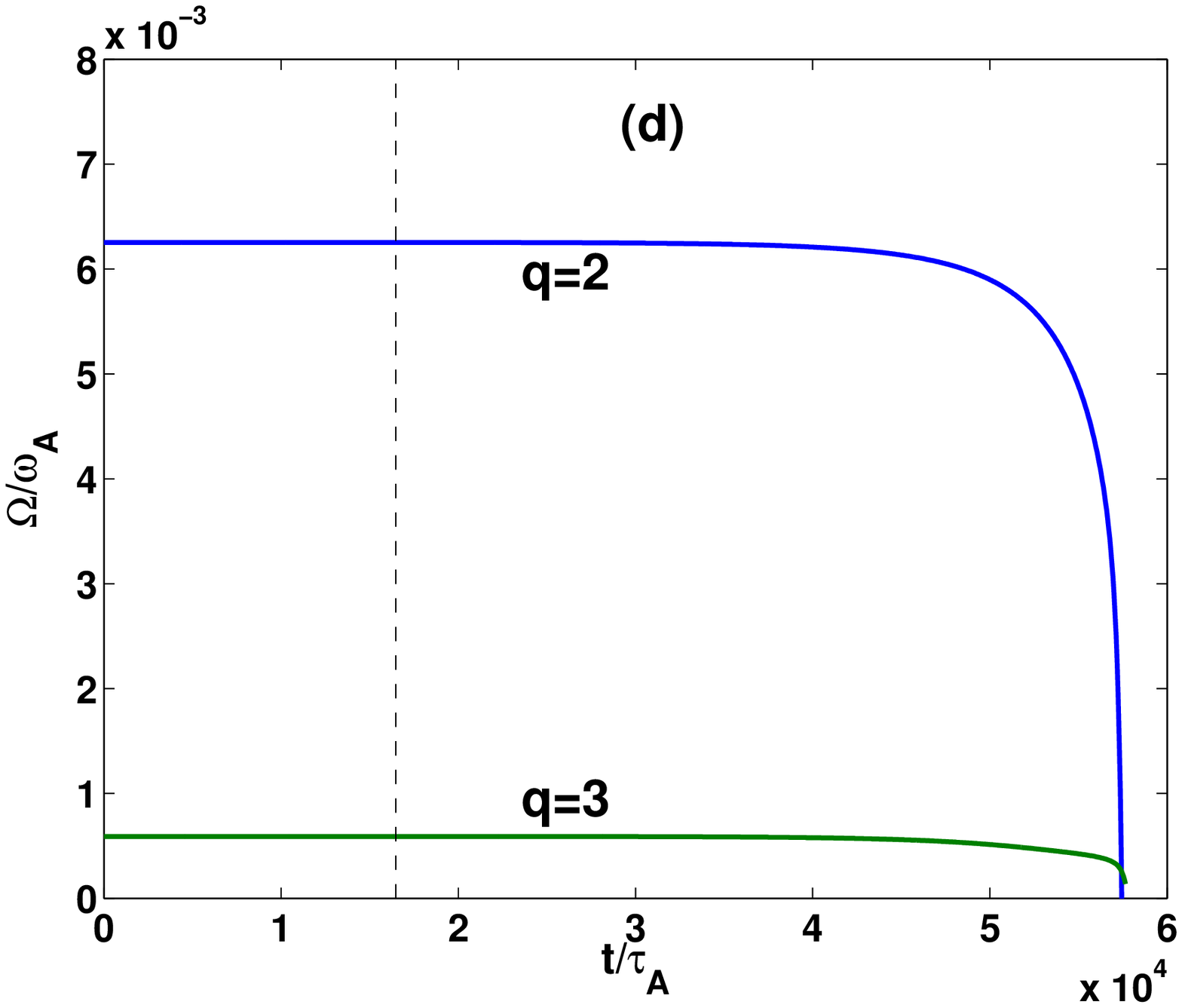}
\caption{Evolution of an initially unstable RWM: (a) the amplitude of
the perturbed radial field $b^1$ at 
the $q=2$ surface, (b) the real and imaginary parts of the perturbed
radial field $b^1$ at the $q=2$ surface, (c) the radial profile of the
plasma rotation frequency, and (d) the plasma rotation frequency at
the $q=2$ and $q=3$ surfaces. The dashed vertical lines in (a,b,d)
indicate the moment of time when the non-linear coupling between the
mode and the plasma flow is switched on. The dashed vertical lines in
(c) indicate the location of the $q=2$ and 3 rational surfaces,
respectively. The numbered lines in (c) correspond to time: 1 -
5.77$\time10^4\tau_A$, 2 - 5.67$\times10^4\tau_A$, and 3 -
5.58$\times10^4\tau_A$. Only the electromagnetic is
included in this simulation.}  
\label{fig:qluj}
\end{center}
\end{figure}

So far most of the theory
\cite{FitzpatrickPP02,GimblettPP04,FitzpatrickPP07} on the rotational
damping due to the onset 
of the RWM has only included the electromagnetic torque as the
momentum sink. We perform a similar simulation by excluding the NTV
torque from the momentum balance equation. The results are summarized
in Fig. \ref{fig:qluj} for the unstable RWM case at
$\Omega_0(t=0)=2.5\times10^{-2}\omega_A$. The electromagnetic torque
alone is also capable of braking the flow, but the effect is generally
more localized near rational surfaces. Similar to the previous case
(see Fig. \ref{fig:qlu}(d)), there is a narrow time window, during which the flow
is rapidly damped, as shown in Fig. \ref{fig:qluj}(d).  

\begin{figure}
\begin{center}
\includegraphics[width=10cm]{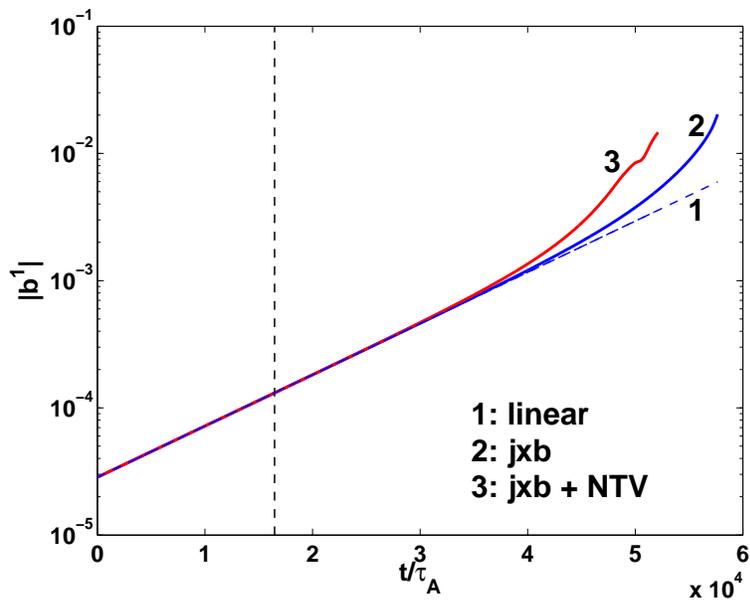}
\caption{Time traces of the amplitude of the perturbed radial
field $b^1$ at the $q=2$ surface, with and without inclusion of the
NTV torque in the simulation. The dashed curve corresponds to the
exponential growth of the initially unstable linear mode. The dashed
vertical line indicates the moment of time when the non-linear
coupling between the mode and the plasma flow is switched on.}  
\label{fig:qlub1}
\end{center}
\end{figure}

In the absence of the NTV torque, the non-linear growth of the RWM
becomes slower during the rotation braking phase, as shown in
Fig. \ref{fig:qlub1}. This is expected taking into account a slower
damping of the flow. 

\subsection{Interaction between initially stable RWM and flow}
A perhaps practically more important problem is the interaction between
a (marginally) stable RWM and the plasma flow. It has been shown, in
both experiments and theory \cite{ChuPPCF10}, that the RWM often
stays marginally stable even in the presence of a strong damping,
either from the kinetic damping or other damping mechanisms. One
example is shown in Fig. \ref{fig:stab}(b) with a strong parallel
sound wave damping. A shallowly stable RWM is sensitive to external
field perturbations, often leading to the so called resonant field amplification
effect \cite{ReimerdesPP06}, which is turn brakes the plasma flow. We investigate
this phenomenon in this Subsection. In particular, we study how the
flow braking is affected by assuming a different level of the initial
(stable) mode amplitude. We assume an initial rotation frequency of
$\Omega_0(t=0)=2.8\times10^{-2}$, which stabilizes the RWM according
to Fig. \ref{fig:stab}(b). 

\begin{figure}
\begin{center}
\includegraphics[width=6.5cm]{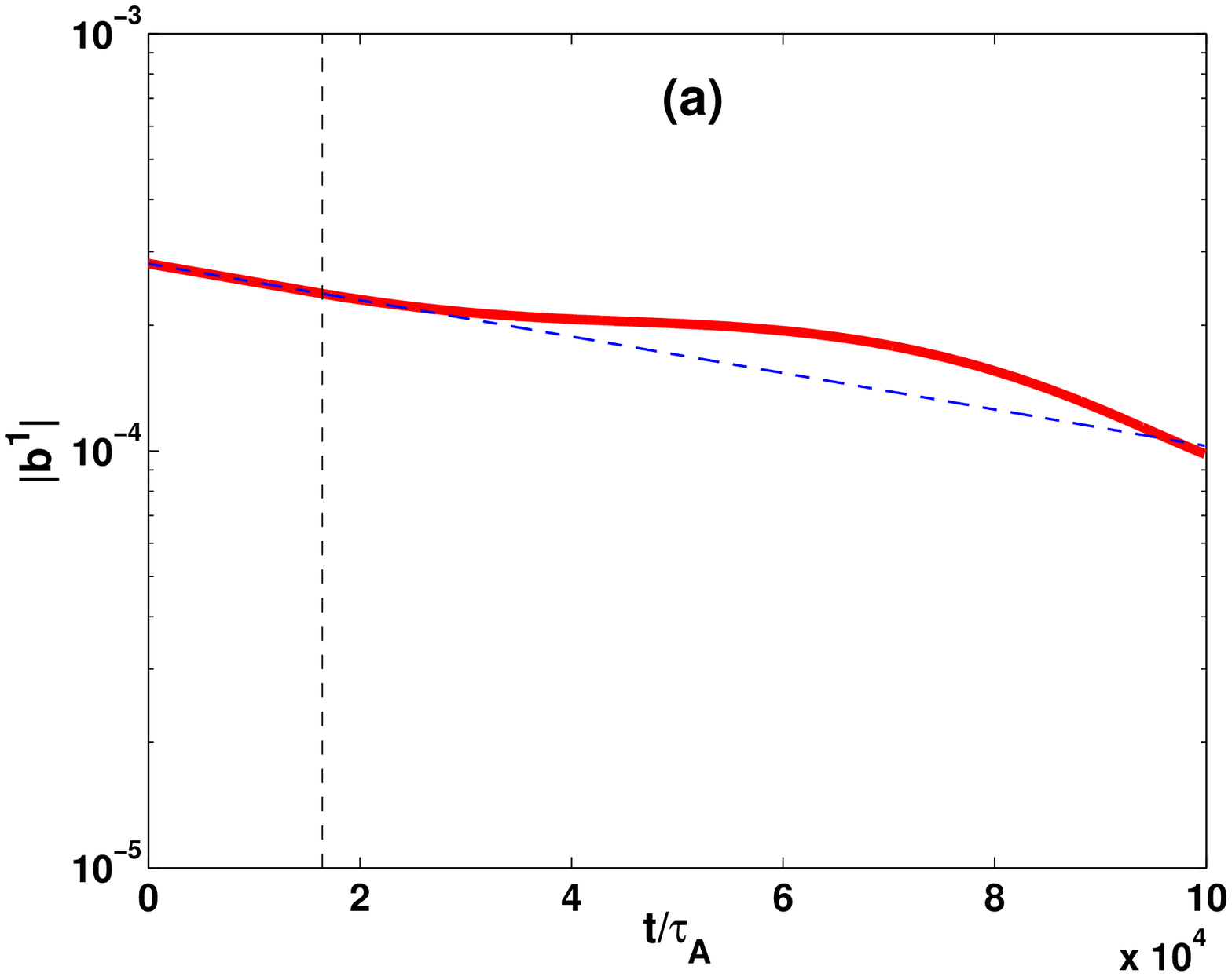}
\includegraphics[width=6.5cm]{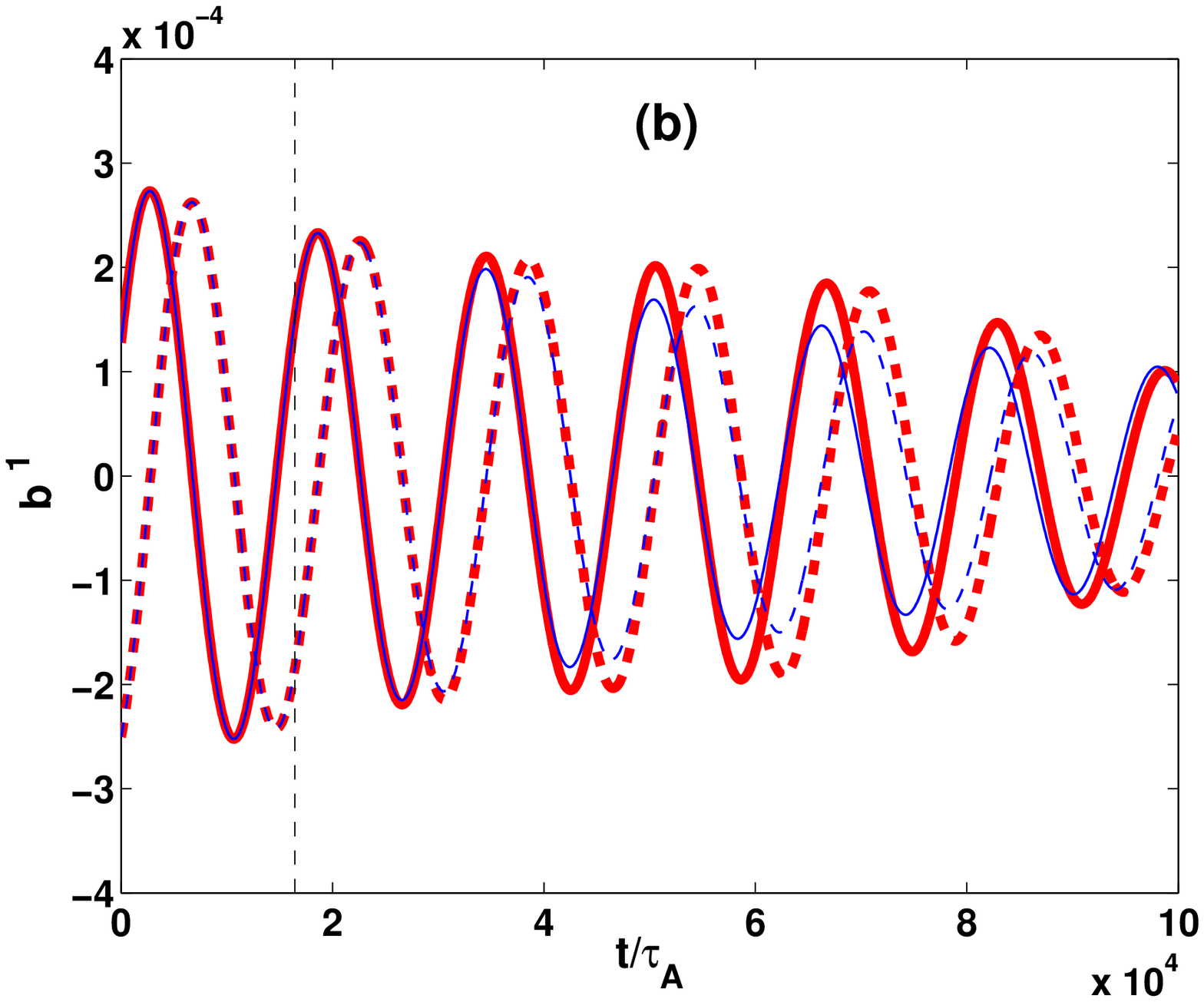}
\includegraphics[width=6.5cm]{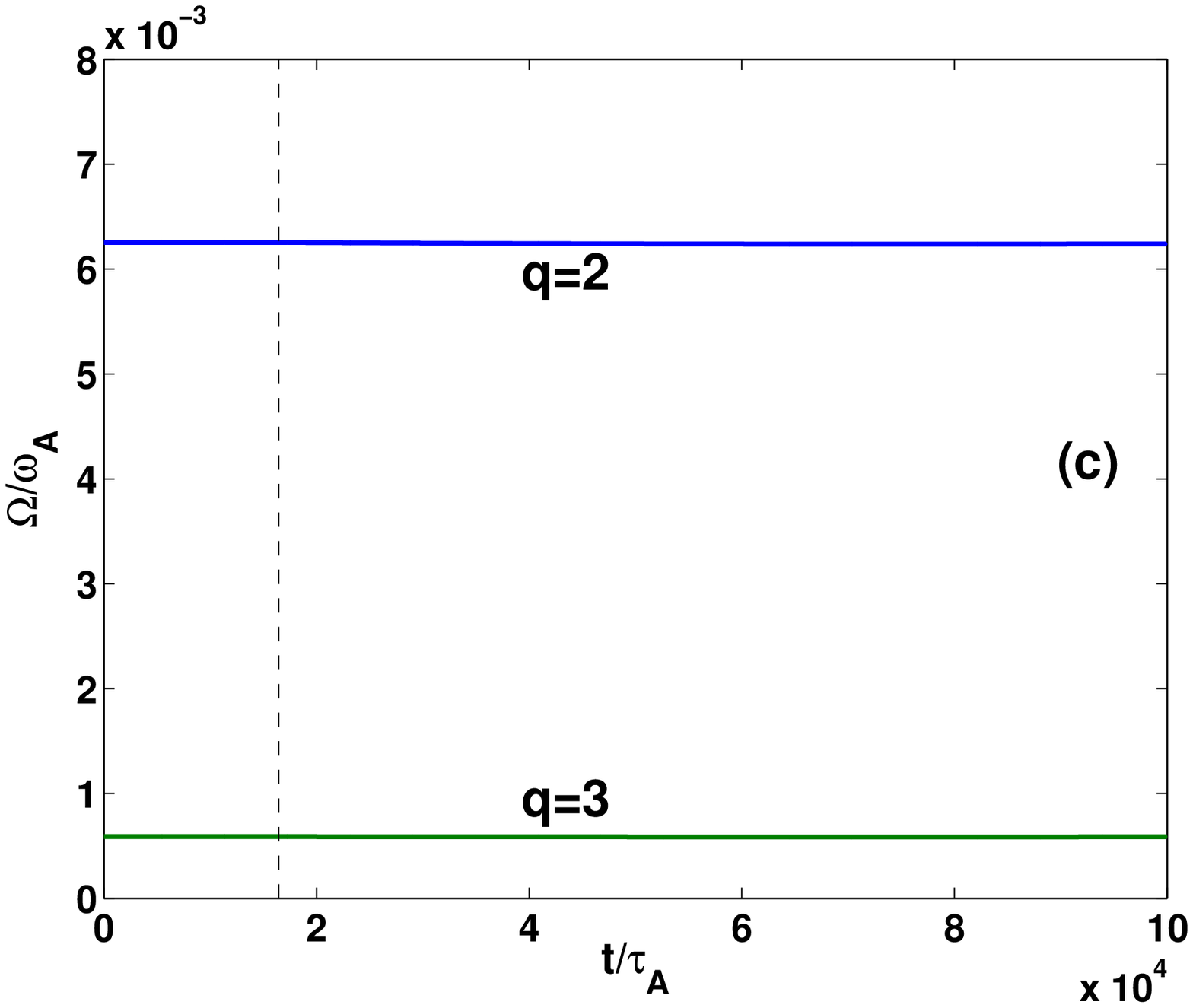}
\includegraphics[width=6.5cm]{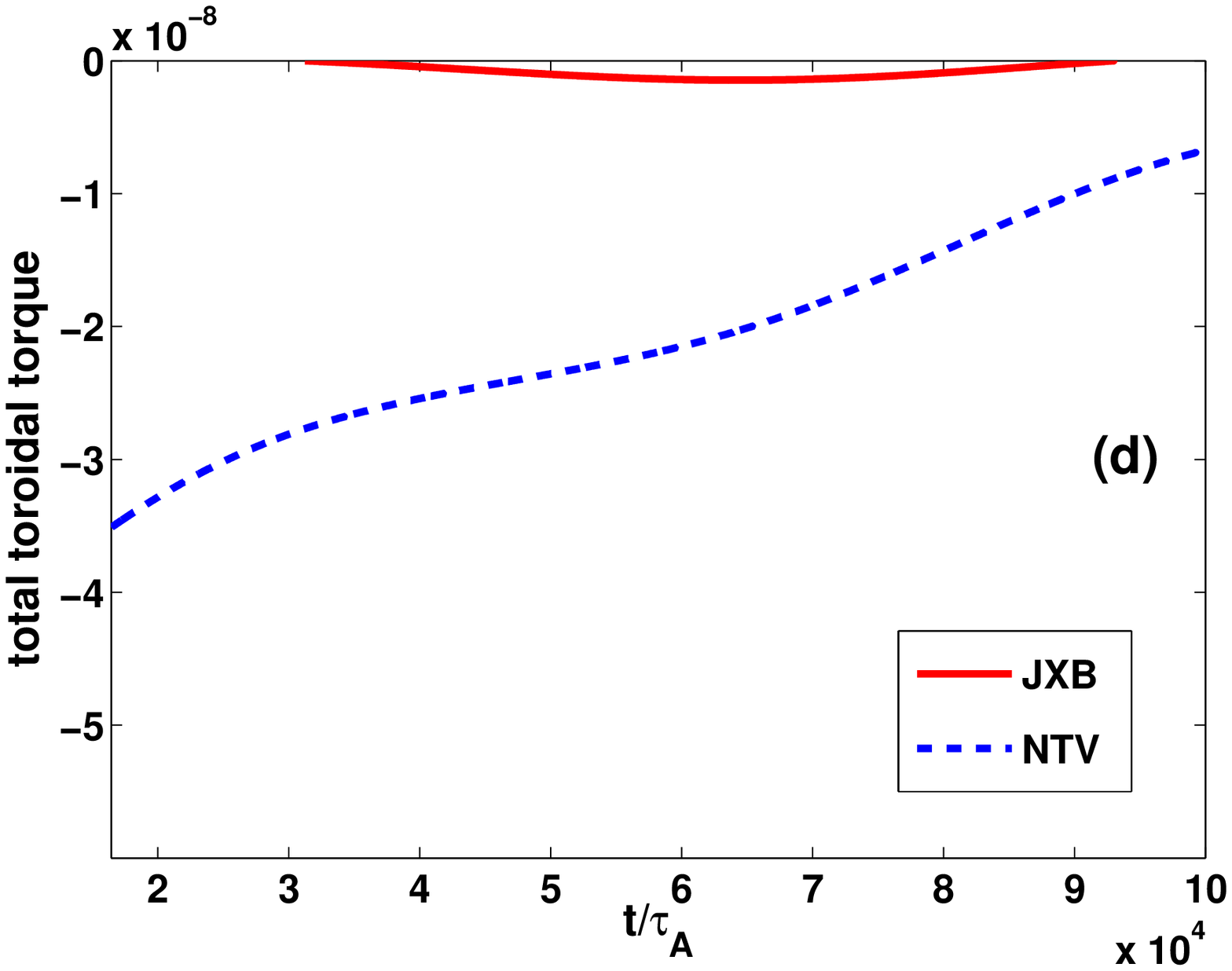}
\caption{Evolution of an initially stable RWM: (a) the amplitude of
the perturbed radial field $b^1$ at 
the $q=2$ surface, (b) the real and imaginary parts of the perturbed
radial field $b^1$ at the $q=2$ surface, (c) the plasma rotation frequency at
the $q=2$ and $q=3$ surfaces, and (d) the net toroidal electromagnetic
and NTV torques acting on the plasma. The dashed vertical lines in (a,b,c)
indicate the moment of time when the non-linear coupling between the
mode and the plasma flow is switched on. The initial mode amplitude,
normalized by $B_0$, is $|b^1|(q=2)=2.8\times10^{-4}$.}  
\label{fig:qls1em3}
\end{center}
\end{figure}

Figure \ref{fig:qls1em3} reports a first case, where we assume a very small
initial mode amplitude,  $|b^1|(q=2,t=0)=2.8\times10^{-4}$. Again the
linear phase is well recovered by the initial value solver, during
$t\in[0,T_1=1.65\times10^4\tau_A]$. When the non-linear coupling
between the mode and the plasma flow is introduced at time $T_1$, no
appreciable braking of the flow is observed, in the presence of both
electromagnetic and NTV torques.       

\begin{figure}
\begin{center}
\includegraphics[width=6.5cm]{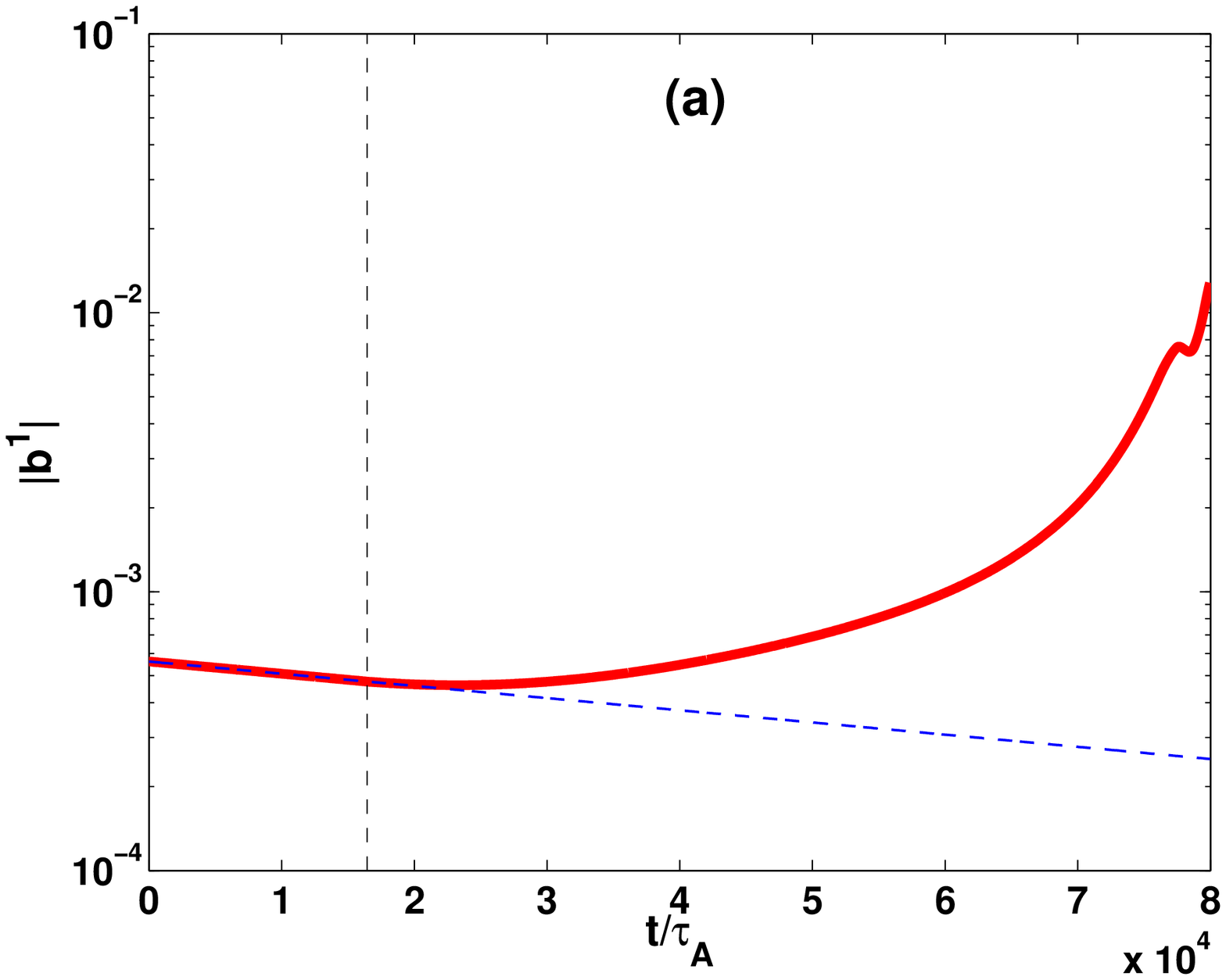}
\includegraphics[width=6.5cm]{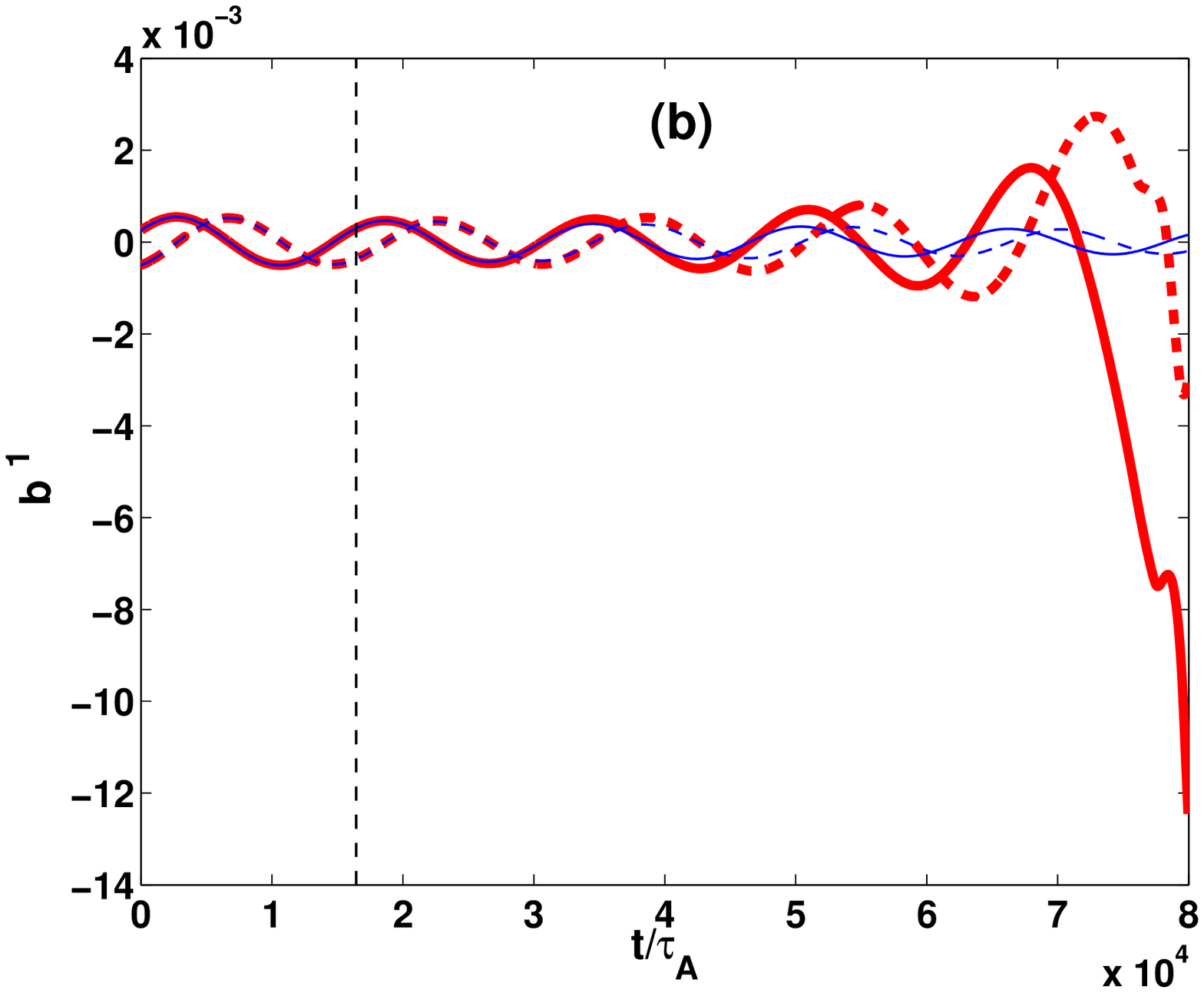}
\includegraphics[width=6.5cm]{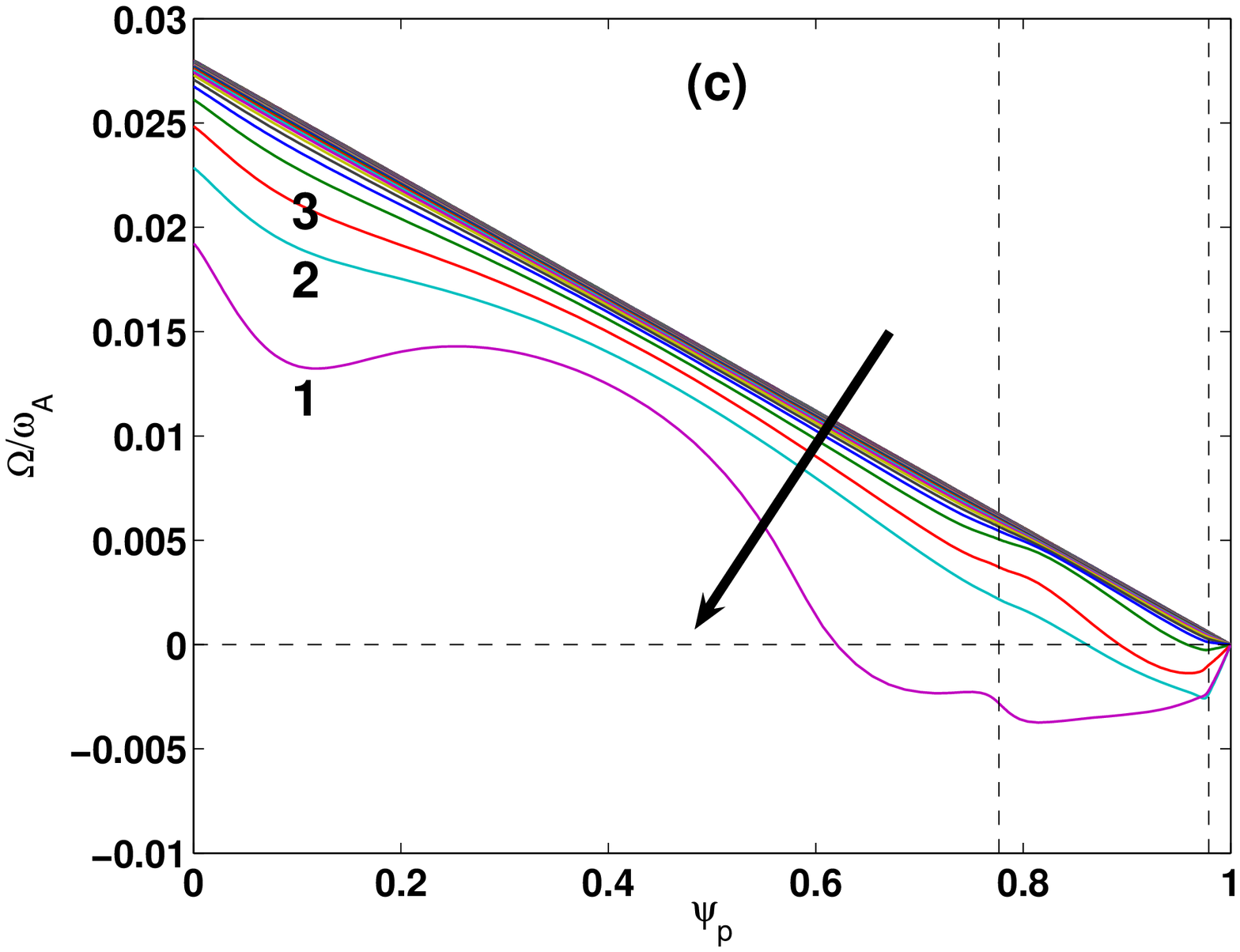}
\includegraphics[width=6.5cm]{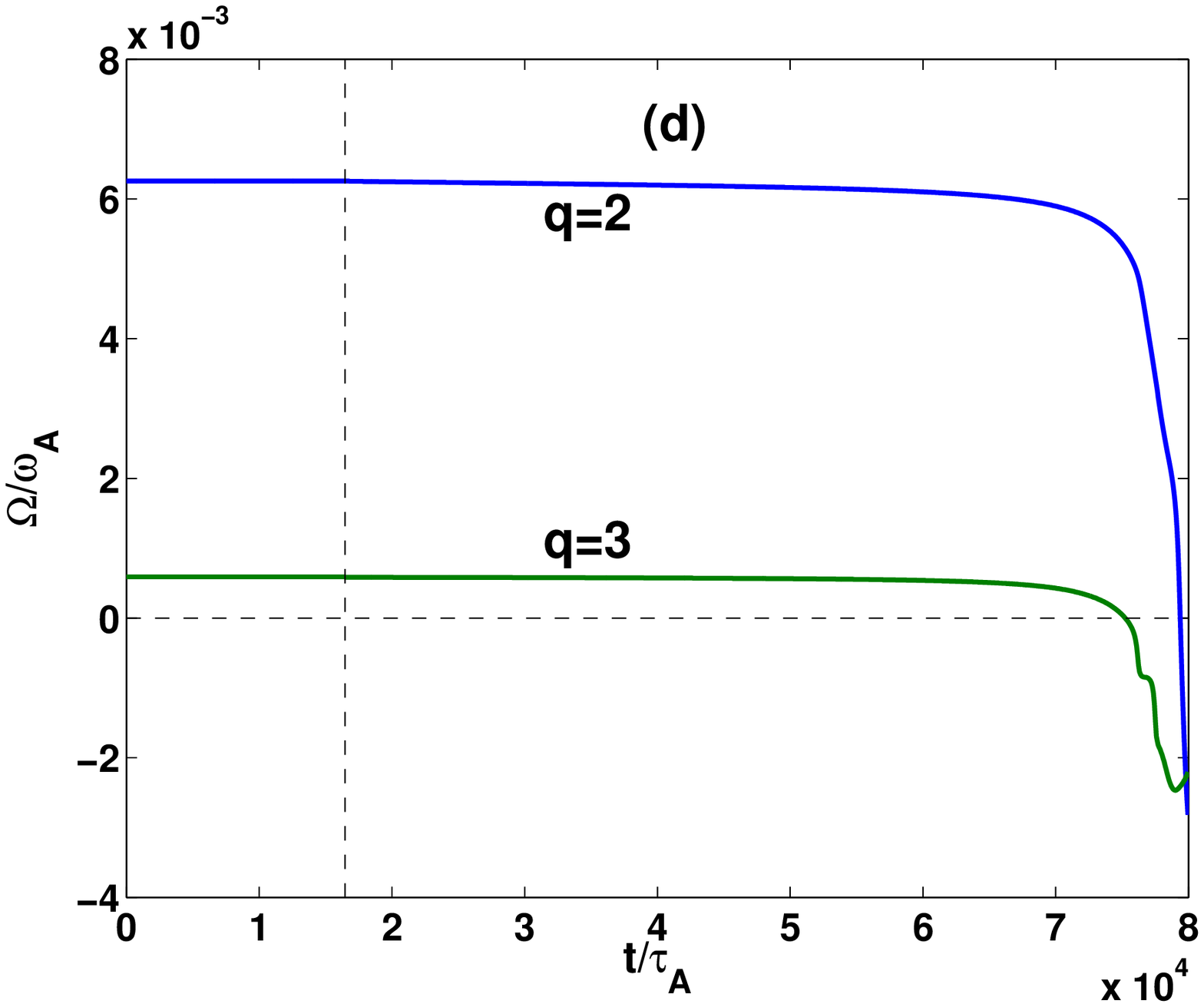}
\caption{Evolution of an initially stable RWM: (a) the amplitude
of the perturbed radial field $b^1$ at 
the $q=2$ surface, (b) the real and imaginary parts of the perturbed
radial field $b^1$ at the $q=2$ surface, (c) the radial profile of the
plasma rotation frequency, and (d) the plasma rotation frequency at
the $q=2$ and $q=3$ surfaces. The dashed vertical lines in (a,b,d)
indicate the moment of time when the non-linear coupling between the
mode and the plasma flow is switched on. The dashed vertical lines in
(c) indicate the location of the $q=2$ and 3 rational surfaces,
respectively. The numbered lines in (c) correspond to time: 1 -
8.00$\time10^4\tau_A$, 2 - 7.86$\times10^4\tau_A$, and 3 -
7.73$\times10^4\tau_A$. The initial mode amplitude,
normalized by $B_0$, is $|b^1|(q=2)=5.6\times10^{-4}$. Both the
electromagnetic and the NTV torques are 
included in this simulation.}  
\label{fig:qls2em3}
\end{center}
\end{figure}

\begin{figure}
\begin{center}
\includegraphics[width=6.5cm]{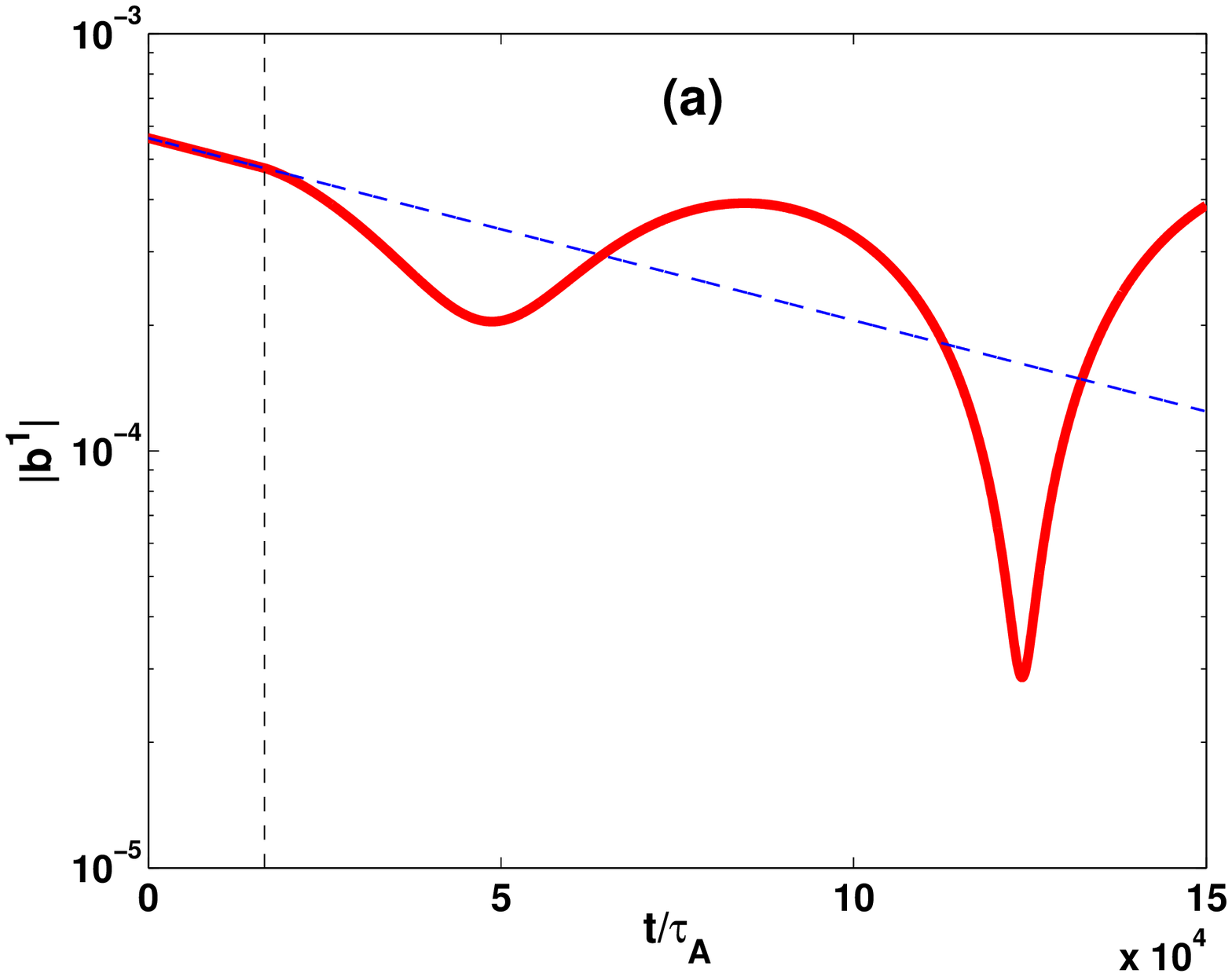}
\includegraphics[width=6.5cm]{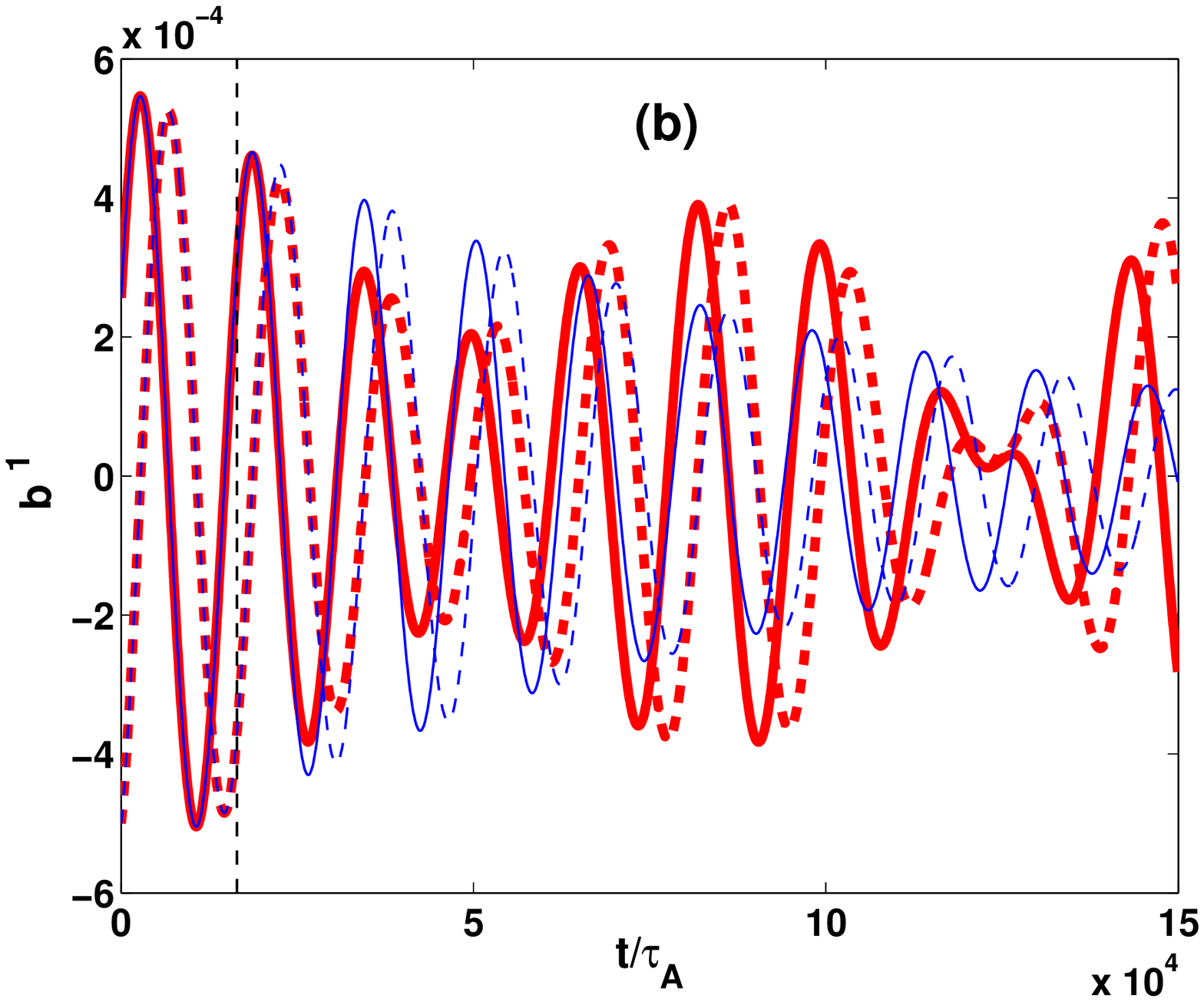}
\includegraphics[width=6.5cm]{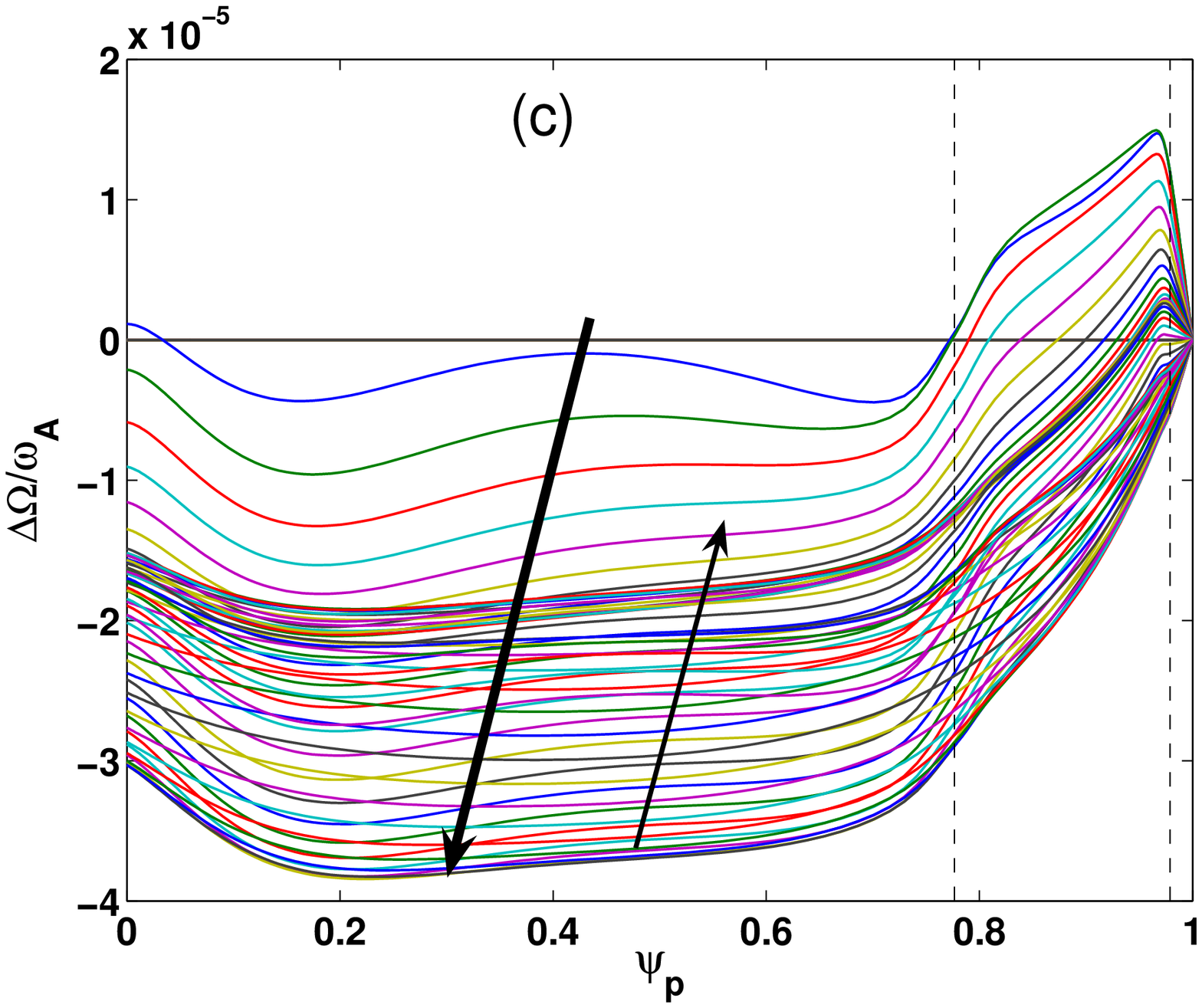}
\includegraphics[width=6.5cm]{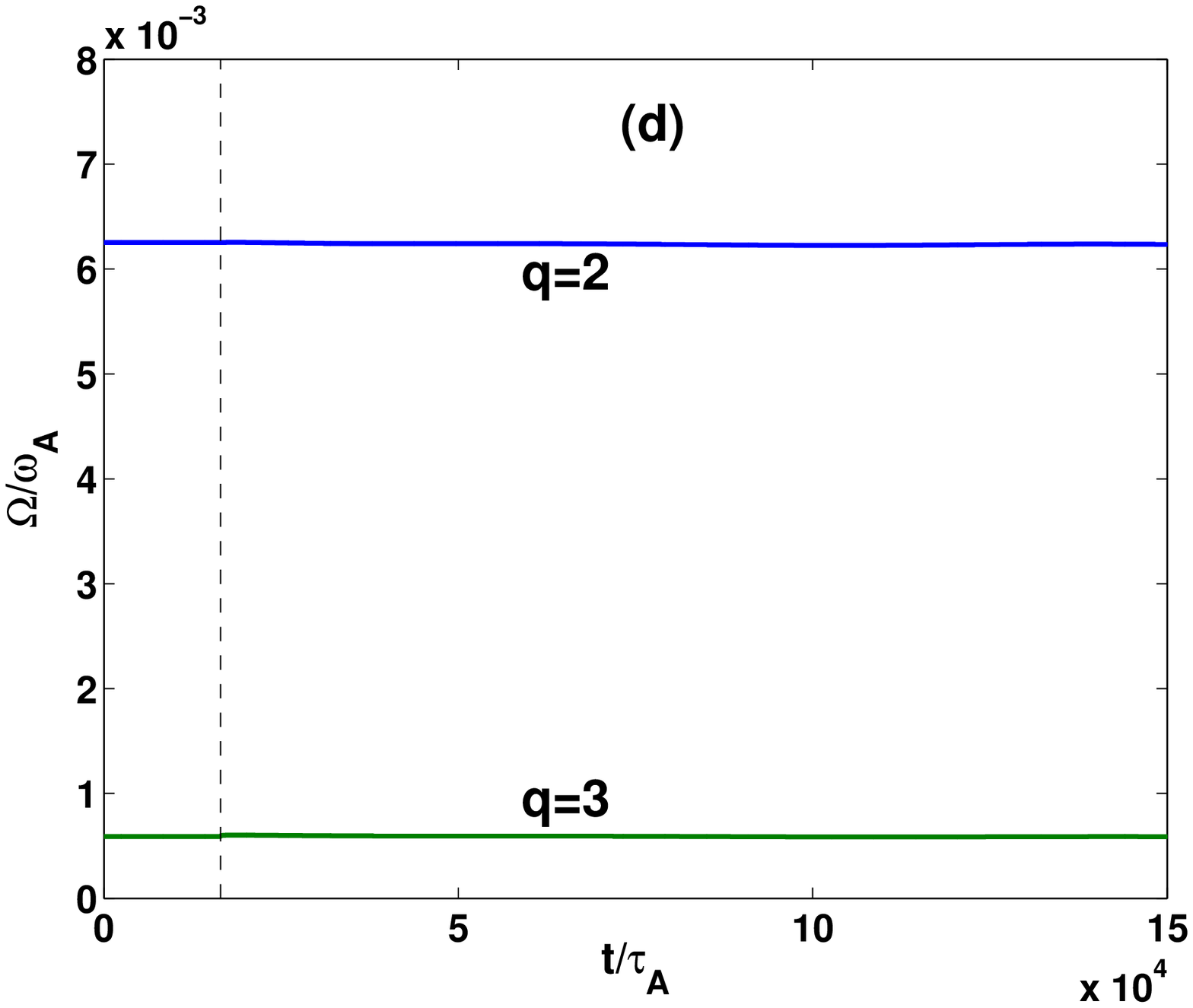}
\caption{Evolution of an initially stable RWM: (a) the amplitude
of the perturbed radial field $b^1$ at 
the $q=2$ surface, (b) the real and imaginary parts of the perturbed
radial field $b^1$ at the $q=2$ surface, (c) the radial profile of the
change of the plasma rotation frequency, and (d) the plasma rotation frequency at
the $q=2$ and $q=3$ surfaces. The dashed vertical lines in (a,b,d)
indicate the moment of time when the non-linear coupling between the
mode and the plasma flow is switched on. The dashed vertical lines in
(c) indicate the location of the $q=2$ and 3 rational surfaces,
respectively. The initial mode amplitude,
normalized by $B_0$, is $|b^1|(q=2)=5.6\times10^{-4}$. Only the
electromagnetic torque is included in this simulation.}  
\label{fig:qls2em3j}
\end{center}
\end{figure}

However, increasing the initial mode amplitude by a factor of 2,
$|b^1|(q=2,t=0)=5.6\times10^{-4}$, results in a full braking of the plasma
flow and the eventual onset of an unstable RWM, as demonstrated by
Fig. \ref{fig:qls2em3}. More interestingly, by excluding the
contribution of the NTV torque from the momentum balance equation, and
keeping the same initial mode amplitude,
only a slight braking of the flow is observed, and the mode stays
stable after a rather long simulation time ($T_{\rm
  end}=1.5\times10^5\omega_A$ after 2520 time steps), as shown by
Fig. \ref{fig:qls2em3j}. Note that Fig. \ref{fig:qls2em3j}(c) shows
the change of the rotation frequency $\Delta\Omega$, which is
non-monotonic, with the radial profile evolution shown by two
arrows. The change is small
compared to the initial rotation frequency. But nevertheless, this
small change of rotation, in particular its radial profile, does
impact the mode evolution as shown in
Fig. \ref{fig:qls2em3j}(a-b). The global radial distribution of
$\Delta\Omega$, at later stage of the simulation, is achieved due to
the momentum diffusion. The faster-than-linear decrease of the
mode amplitude, during certain periods of time, is associated with the
change of the radial profile of flow, and consequently, the deviation of
the mode structure from that of the linear eigenmode.      

\begin{figure}
\begin{center}
\includegraphics[width=6.5cm]{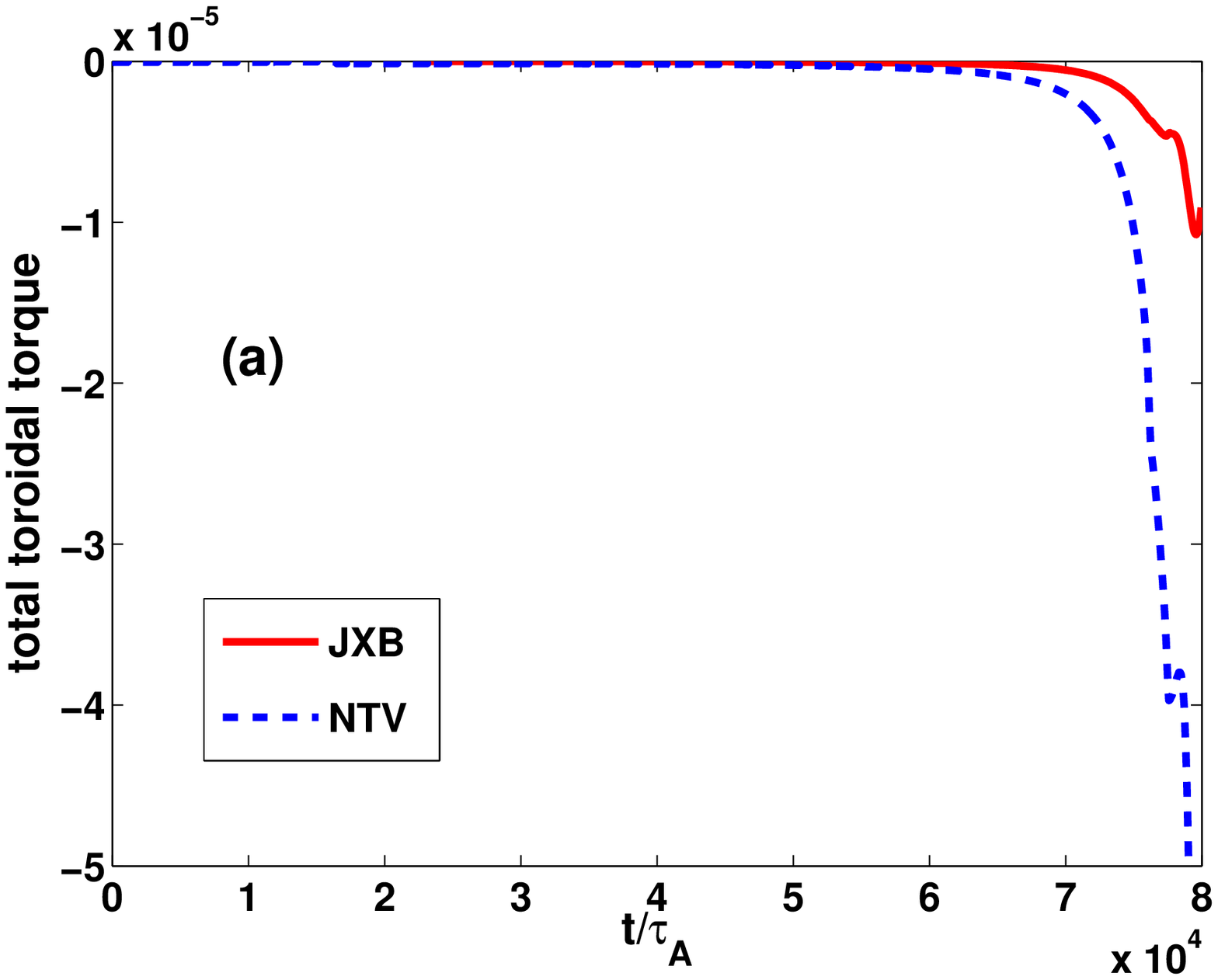}
\includegraphics[width=6.5cm]{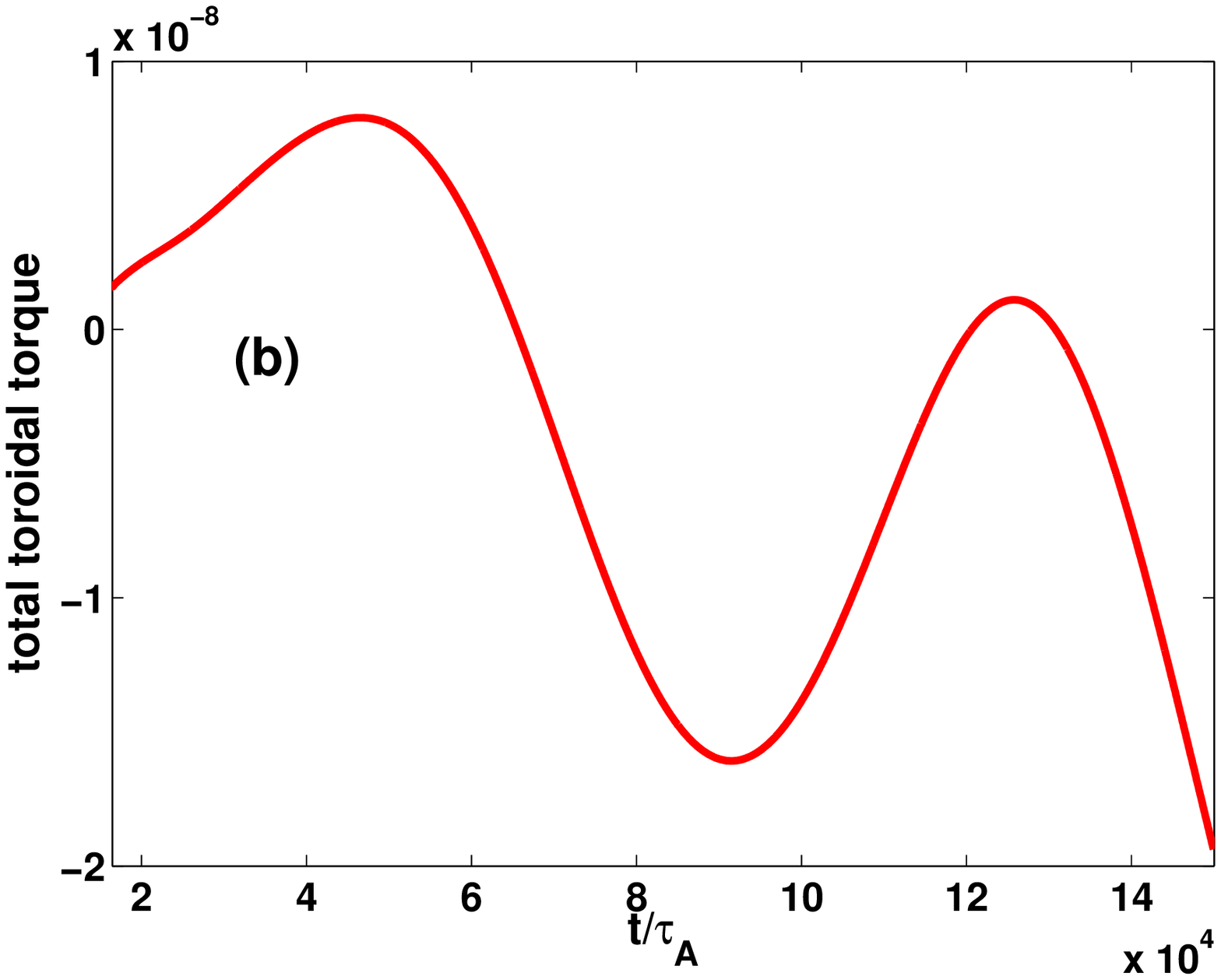}
\caption{Time traces of the net toroidal electromagnetic (solid line)
and NTV (dashed line) torques
acting on the plasma, during the non-linear evolution of an initially
stable RWM, with the initial mode amplitude
$|b^1|(q=2)=5.6\times10^{-4}$: (a) both electromagnetic and NTV
torques are included in the toroidal torque balance, and (b) only the
electromagnetic torque is included.}  
\label{fig:qlstorq}
\end{center}
\end{figure}

A comparison of the net electromagnetic versus NTV torque, shown in
Fig. \ref{fig:qlstorq}(a), again confirms that the latter contributes
the dominant momentum sink. In the absence of the NTV torque
(Fig. \ref{fig:qlstorq}(b)), the electromagnetic torque is too small
to give an appreciable effect on the flow damping.   

\begin{figure}
\begin{center}
\includegraphics[width=6.5cm]{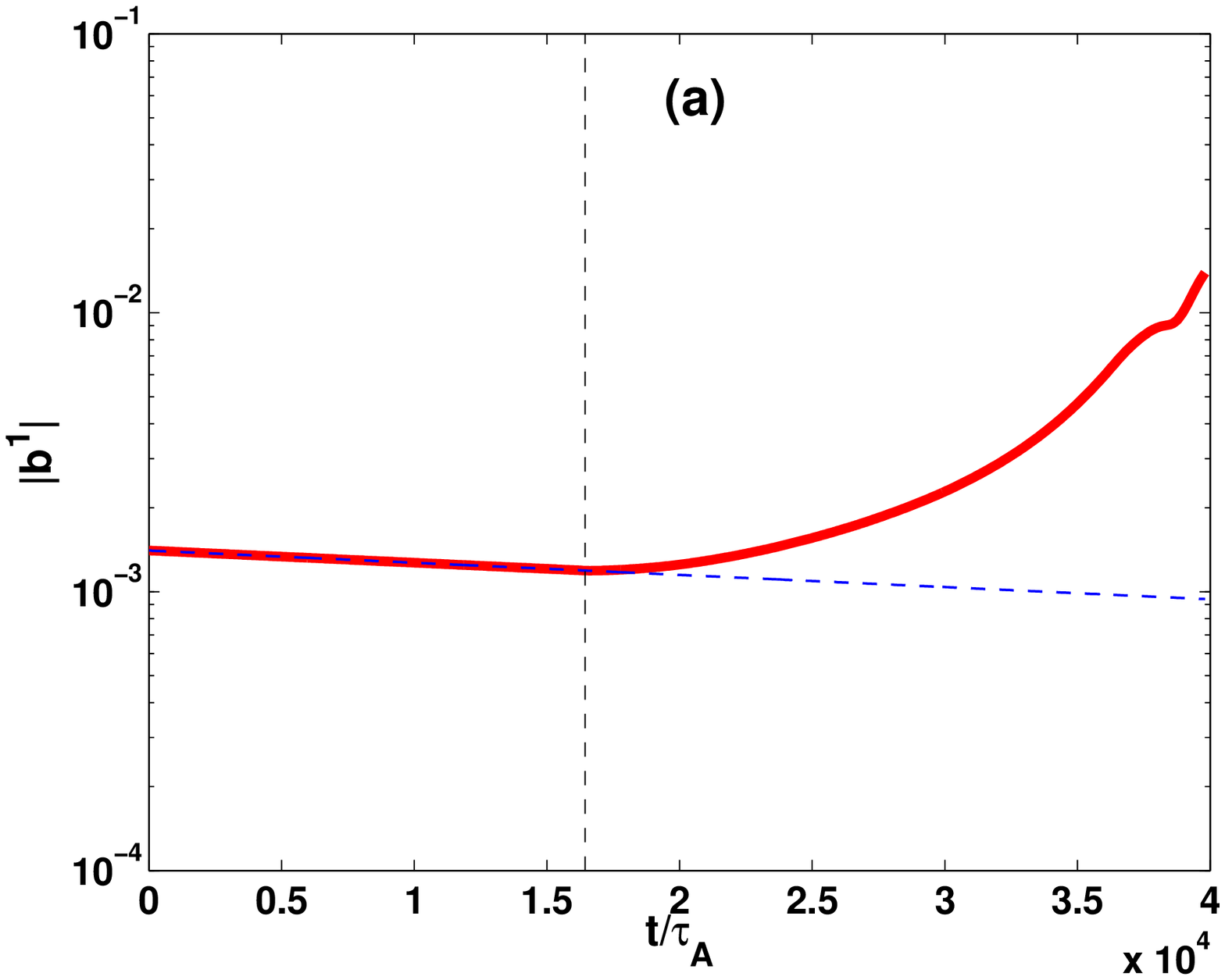}
\includegraphics[width=6.5cm]{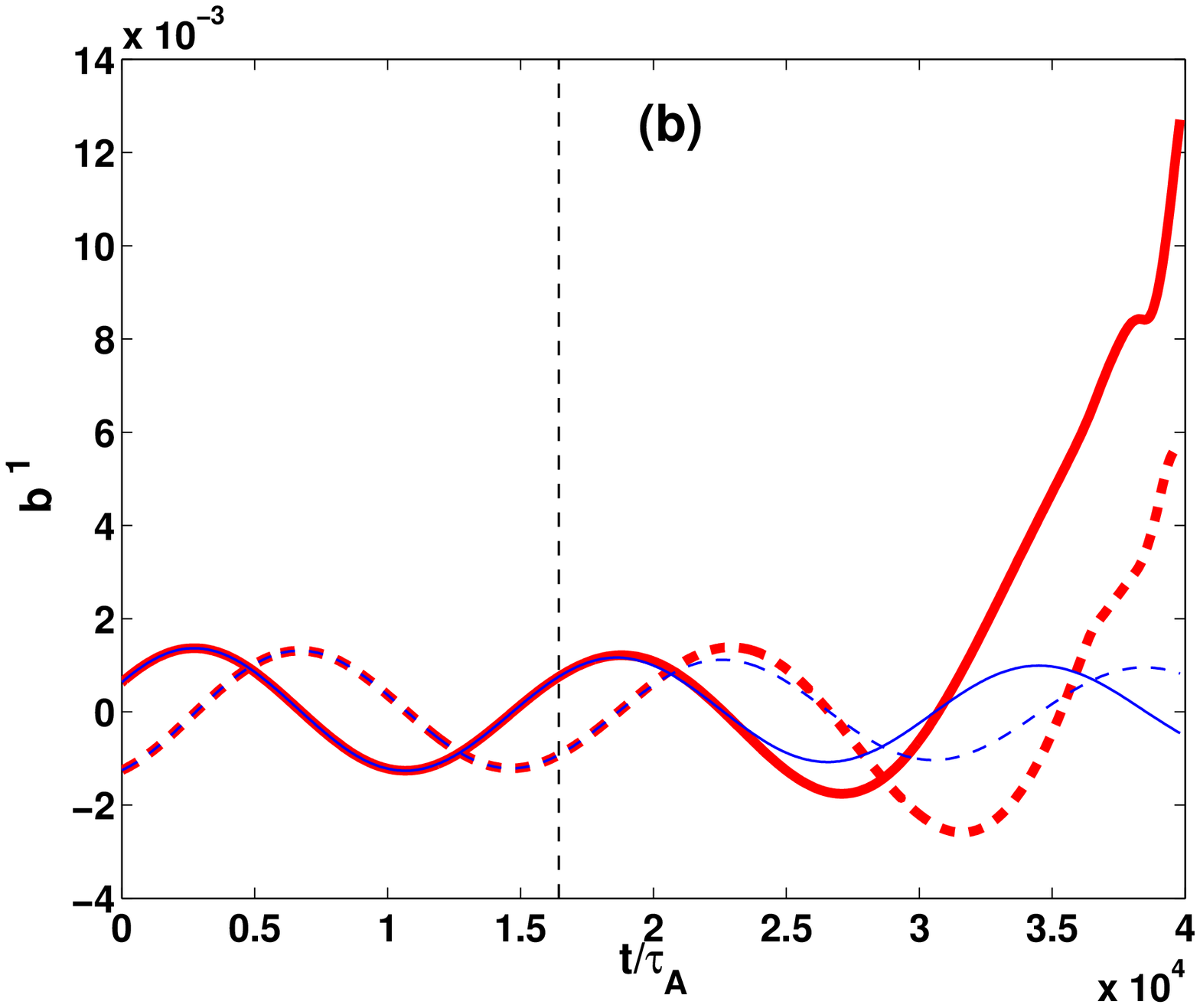}
\includegraphics[width=6.5cm]{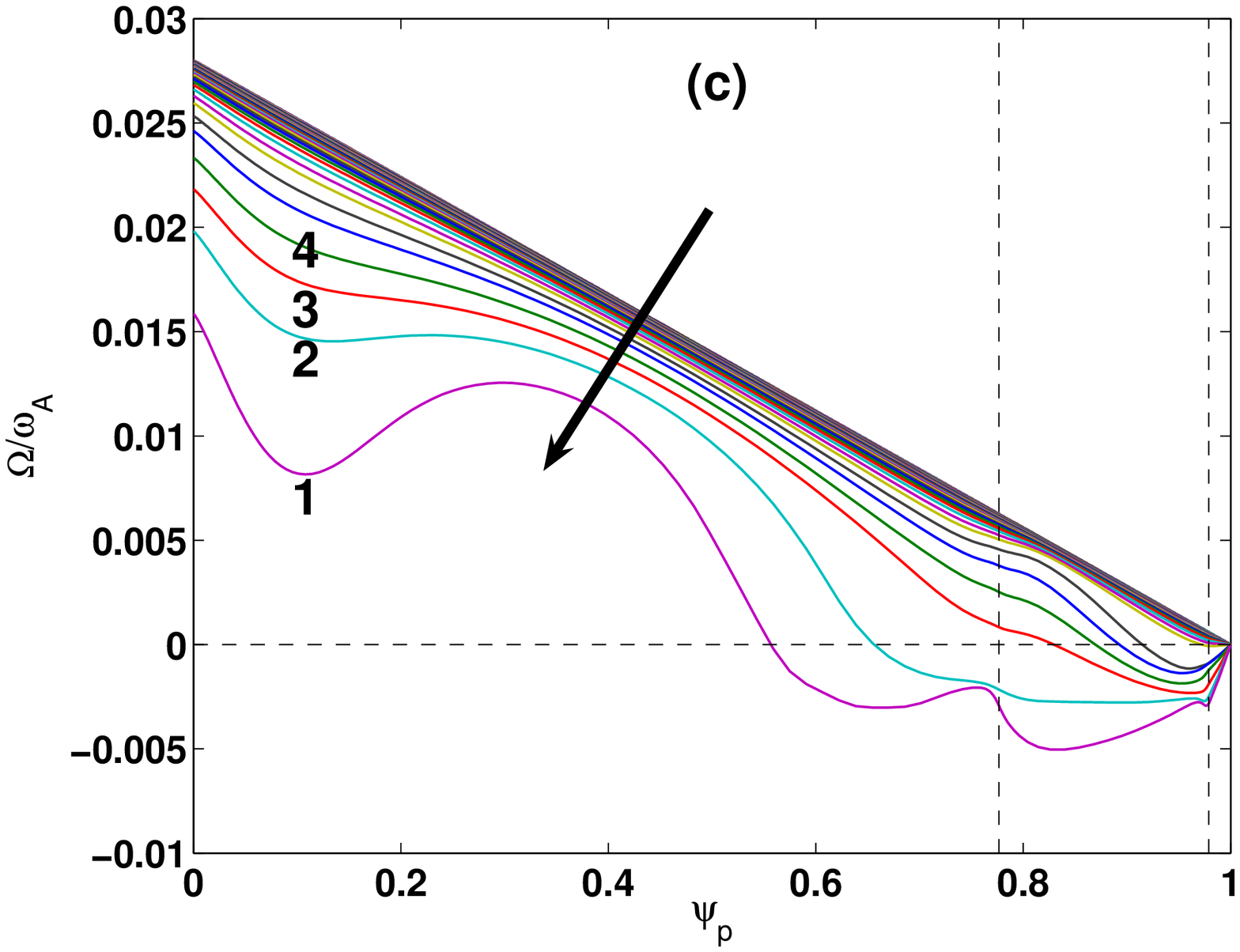}
\includegraphics[width=6.5cm]{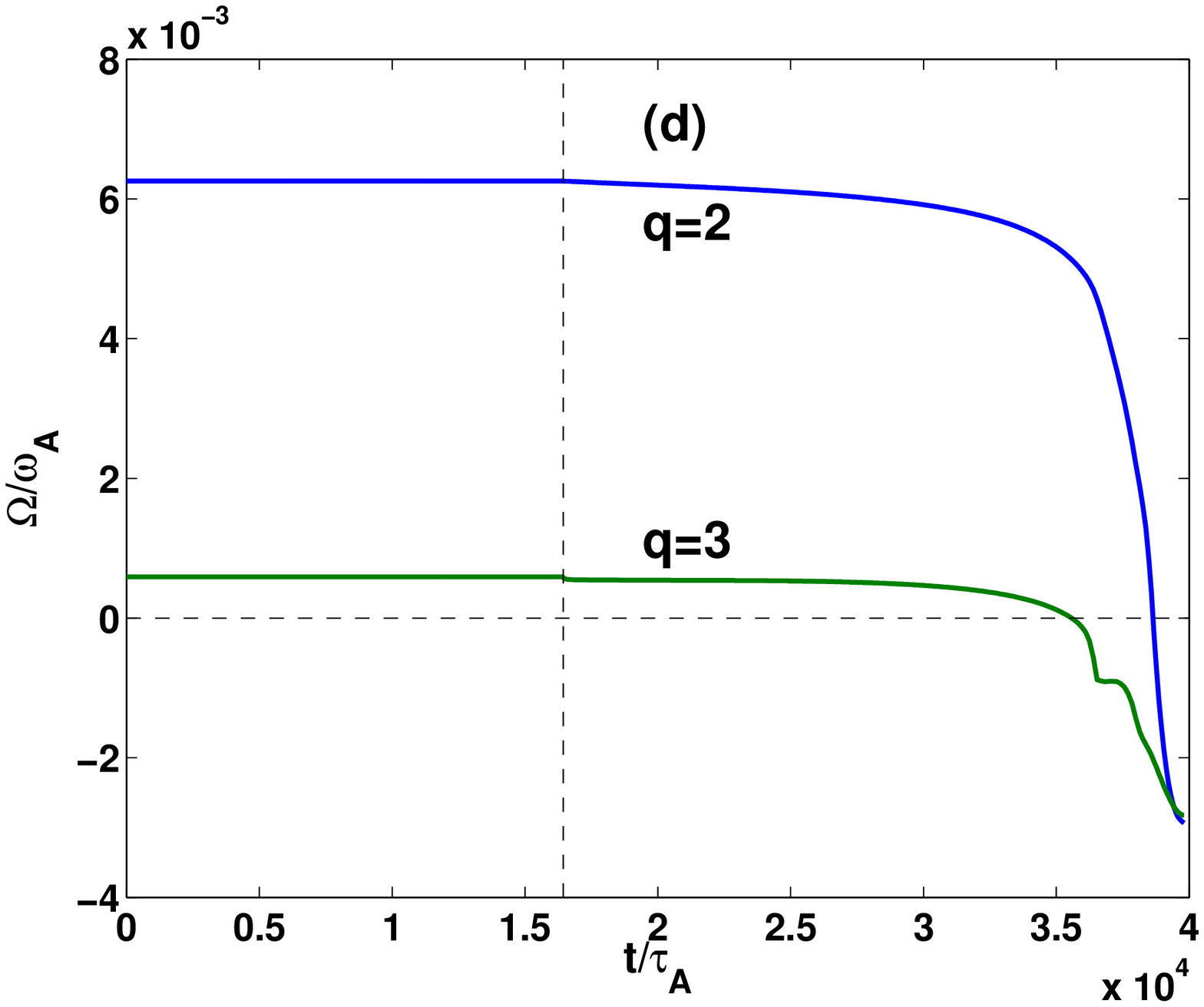}
\caption{Evolution of an initially stable RWM: (a) the amplitude
of the perturbed radial field $b^1$ at 
the $q=2$ surface, (b) the real and imaginary parts of the perturbed
radial field $b^1$ at the $q=2$ surface, (c) the radial profile of the
plasma rotation frequency, and (d) the plasma rotation frequency at
the $q=2$ and $q=3$ surfaces. The dashed vertical lines in (a,b,d)
indicate the moment of time when the non-linear coupling between the
mode and the plasma flow is switched on. The dashed vertical lines in
(c) indicate the location of the $q=2$ and 3 rational surfaces,
respectively. The numbered lines in (c) correspond to time: 1 -
3.98$\time10^4\tau_A$, 2 - 3.91$\times10^4\tau_A$, 3 - 3.85$\times10^4\tau_A$, and 4 -
3.79$\times10^4\tau_A$. The initial mode amplitude,
normalized by $B_0$, is $|b^1|(q=2)=1.4\times10^{-3}$. Both the
electromagnetic and the NTV torques are 
included in this simulation.}  
\label{fig:qls5em3}
\end{center}
\end{figure}

\begin{figure}
\begin{center}
\includegraphics[width=6.5cm]{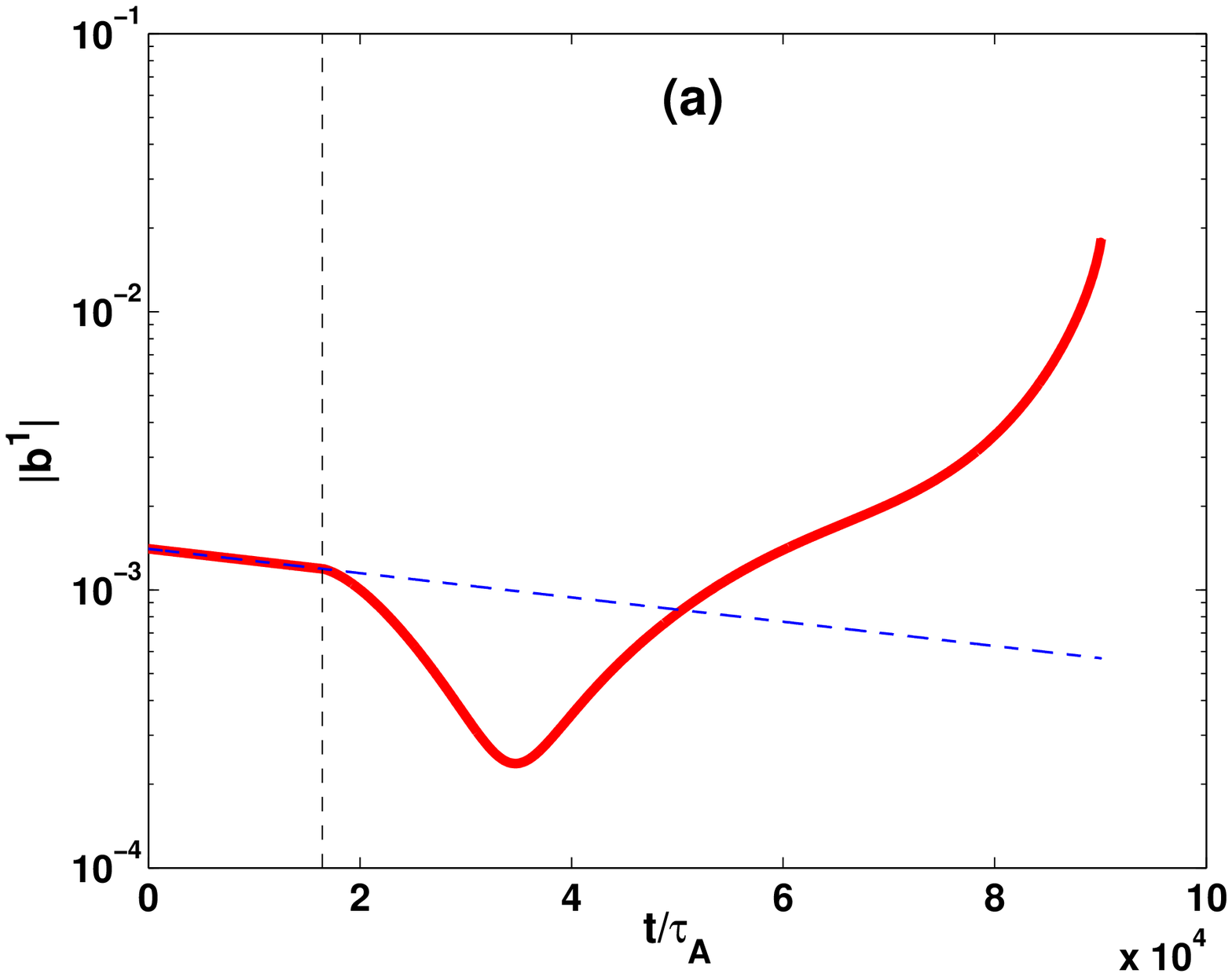}
\includegraphics[width=6.5cm]{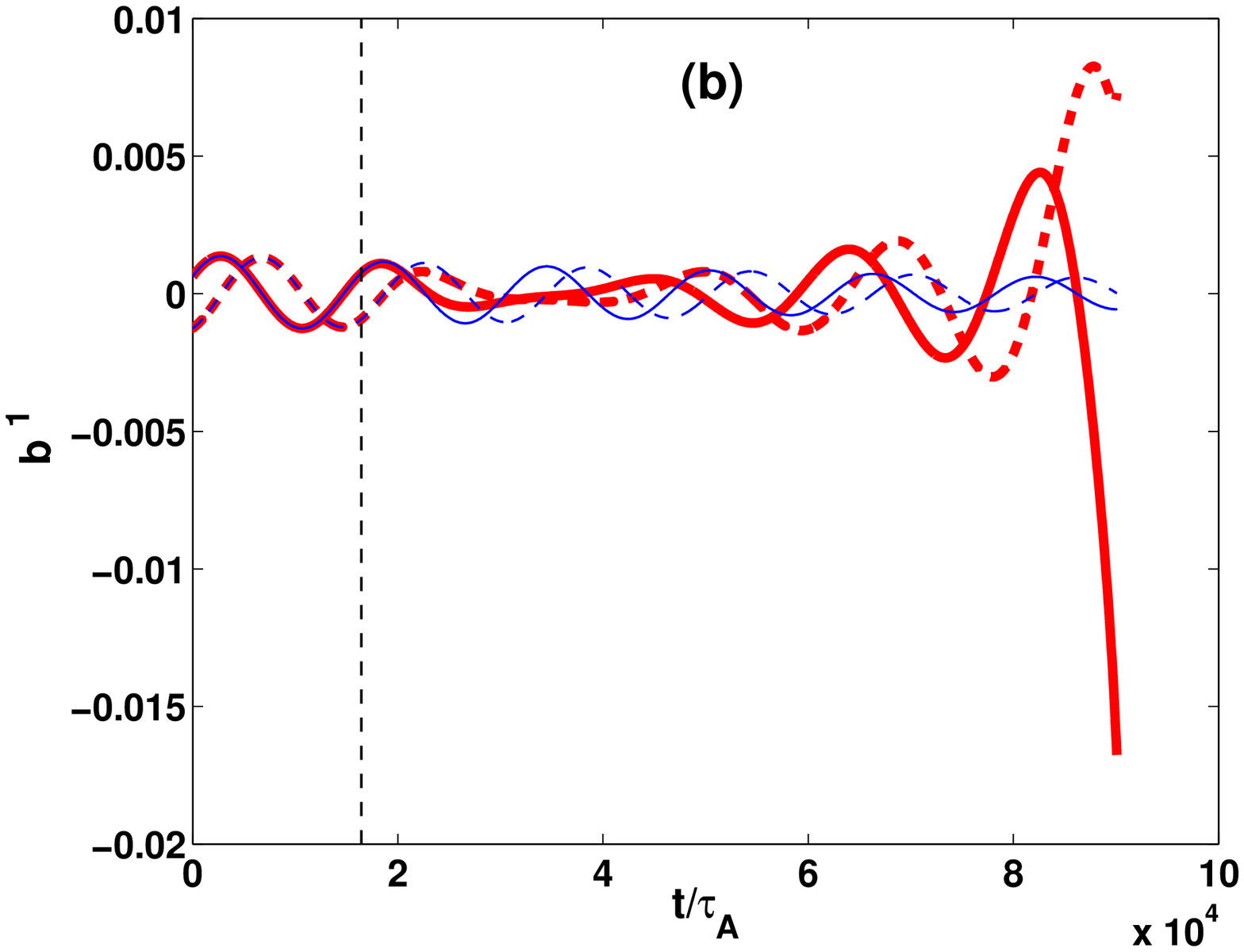}
\includegraphics[width=6.5cm]{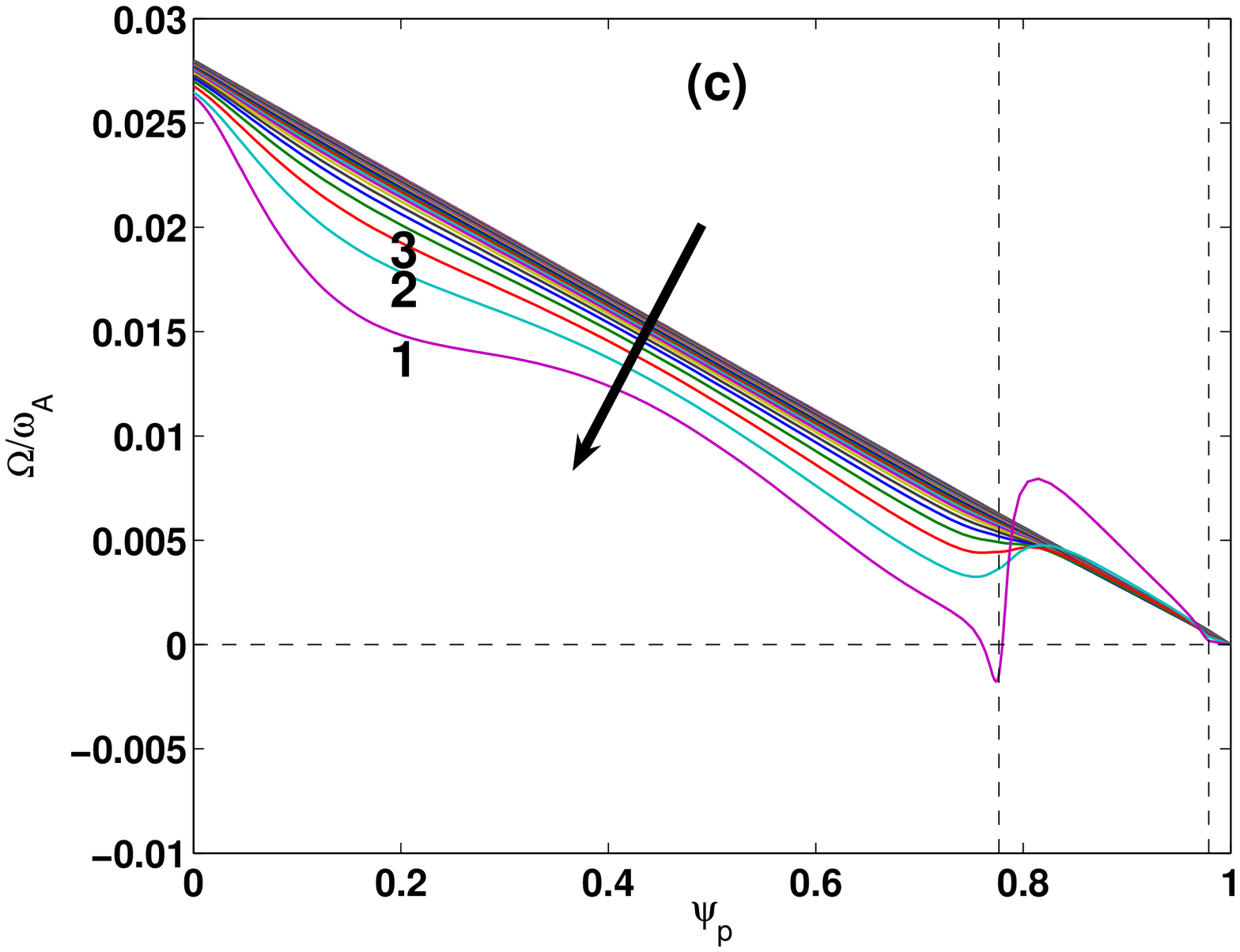}
\includegraphics[width=6.5cm]{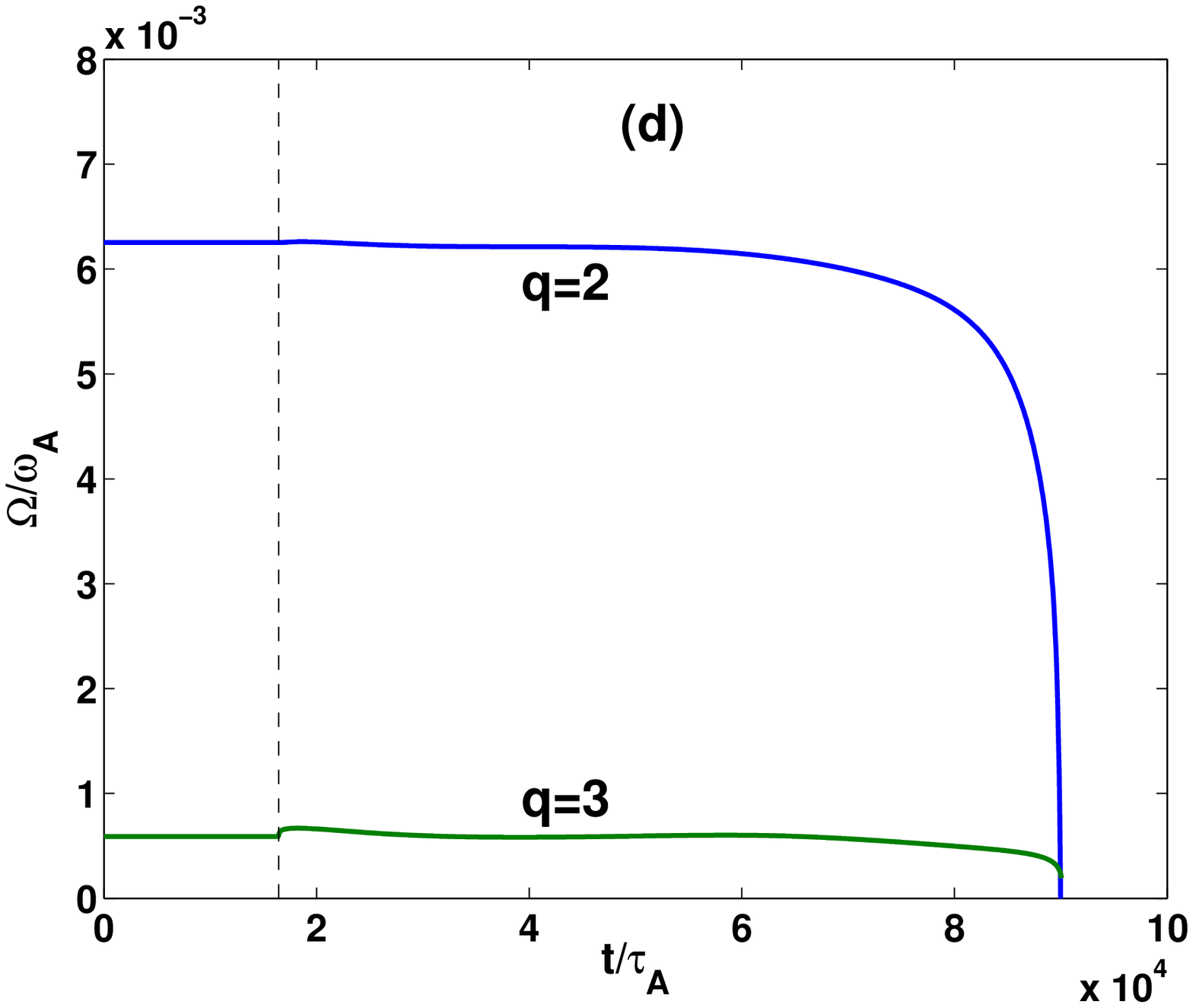}
\caption{Evolution of an initially stable RWM: (a) the amplitude
of the perturbed radial field $b^1$ at 
the $q=2$ surface, (b) the real and imaginary parts of the perturbed
radial field $b^1$ at the $q=2$ surface, (c) the radial profile of the
plasma rotation frequency, and (d) the plasma rotation frequency at
the $q=2$ and $q=3$ surfaces. The dashed vertical lines in (a,b,d)
indicate the moment of time when the non-linear coupling between the
mode and the plasma flow is switched on. The dashed vertical lines in
(c) indicate the location of the $q=2$ and 3 rational surfaces,
respectively. The numbered lines in (c) correspond to time: 1 -
9.00$\time10^4\tau_A$, 2 - 8.85$\times10^4\tau_A$, and 3 -
8.70$\times10^4\tau_A$. The initial mode amplitude,
normalized by $B_0$, is $|b^1|(q=2)=1.4\times10^{-3}$. Only the
electromagnetic torque is included in this simulation.}  
\label{fig:qls5em3j}
\end{center}
\end{figure}

\begin{figure}
\begin{center}
\includegraphics[width=6.5cm]{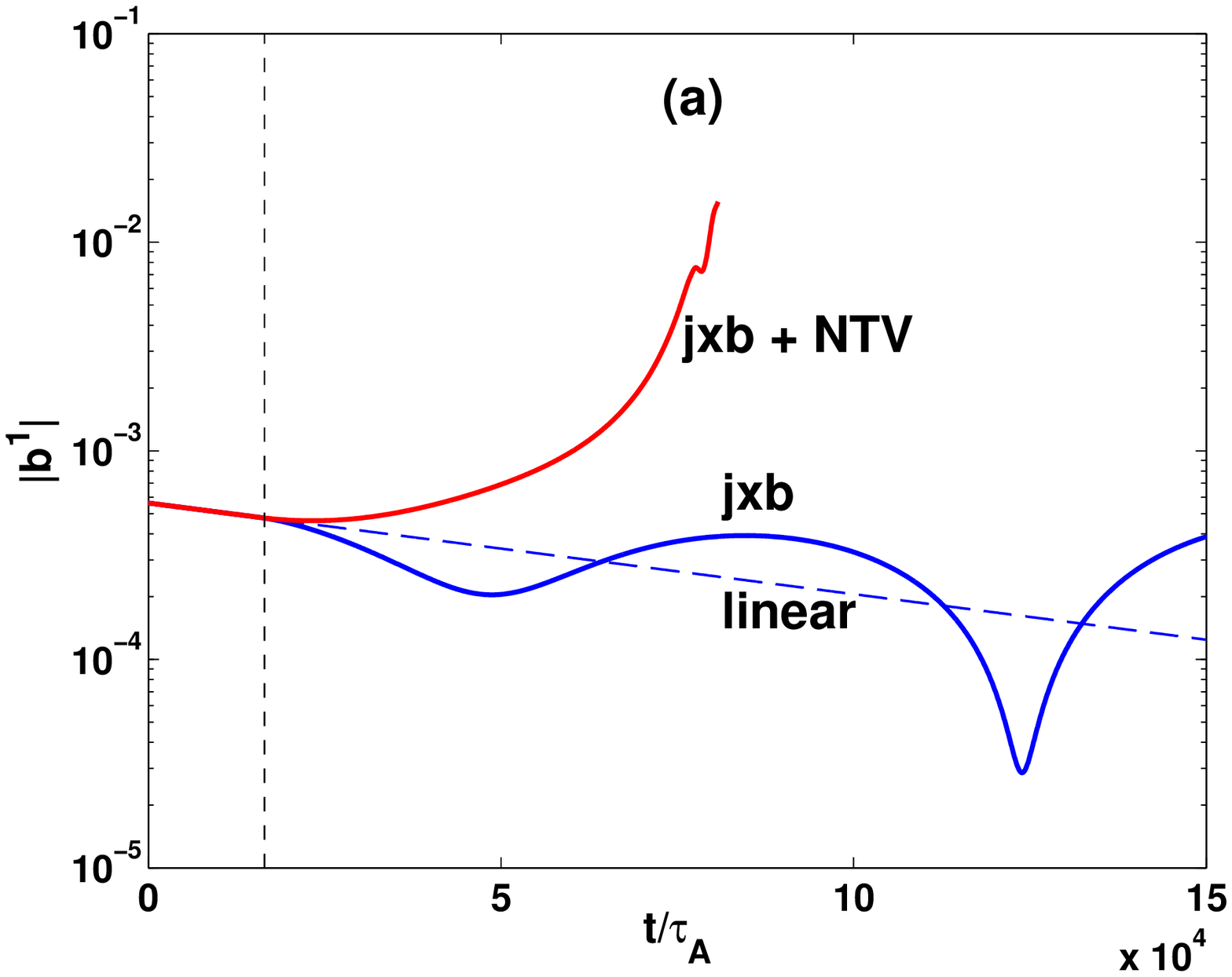}
\includegraphics[width=6.5cm]{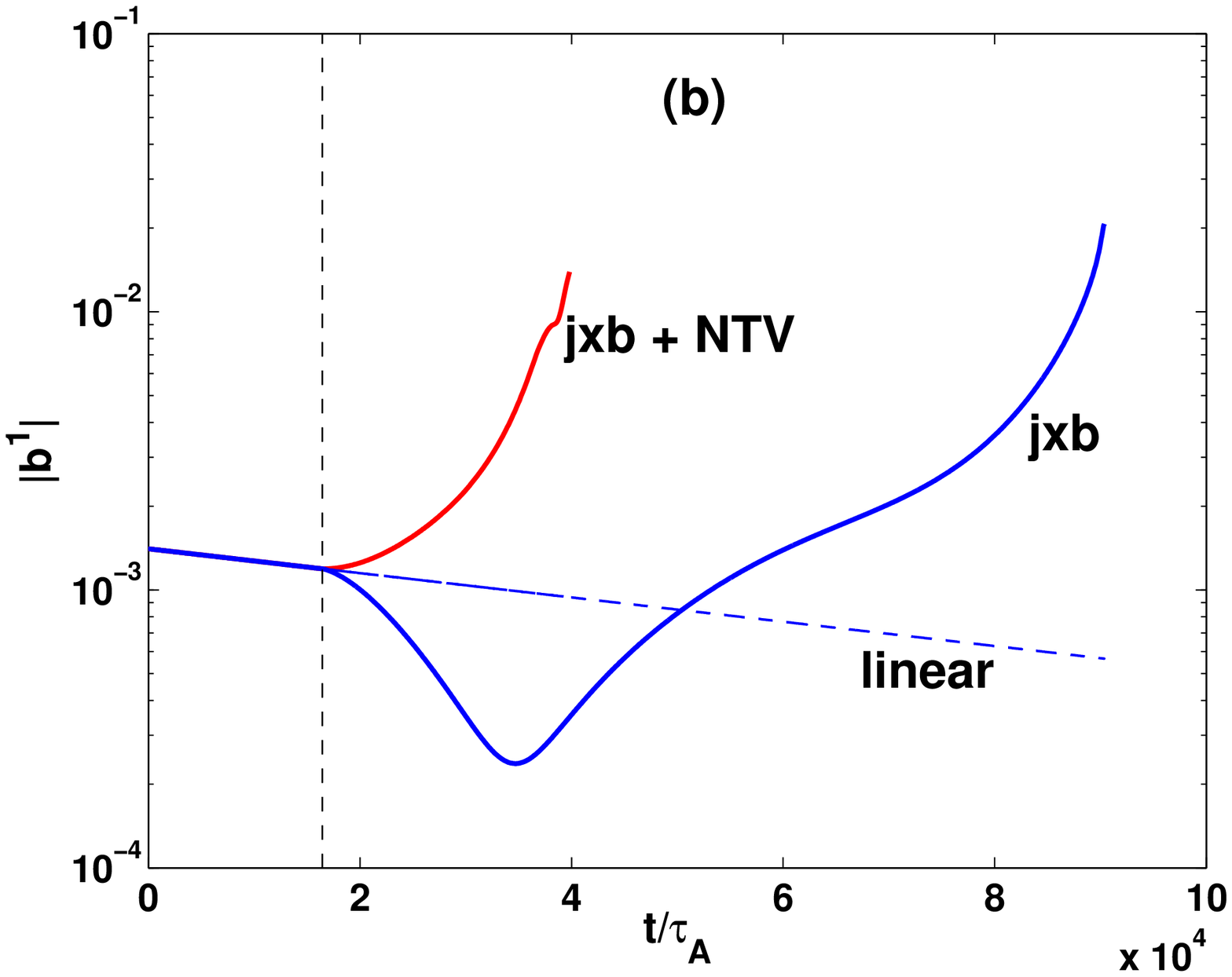}
\caption{Time traces of the amplitude of the perturbed radial
field $b^1$ at the $q=2$ surface, with and without inclusion of the
NTV torque in the simulations for an initially
stable RWM, with the initial mode amplitude at (a)
$|b^1|(q=2)=5.6\times10^{-4}$, and (b) $|b^1|(q=2)=1.4\times10^{-3}$.
The dashed curves correspond to the 
exponential decay of the initially stable linear mode. The dashed
vertical lines indicate the moment of time when the non-linear
coupling between the mode and the plasma flow is switched on.}  
\label{fig:qlsb1}
\end{center}
\end{figure}

Further increase of the initial mode amplitude leads to a full rotation
braking with or without the NTV contributions. Figures
\ref{fig:qls5em3} and \ref{fig:qls5em3j} show one example, where the
initial mode amplitude is increased by 5 times compared to that of
Fig. \ref{fig:qls1em3}.  Figures \ref{fig:qlsb1}(a-b) compare time
traces of the mode amplitude $|b^1|(q=2)$ with or without the NTV
torque, for two different values of the initial mode amplitude.  

\begin{figure}
\begin{center}
\includegraphics[width=6.5cm]{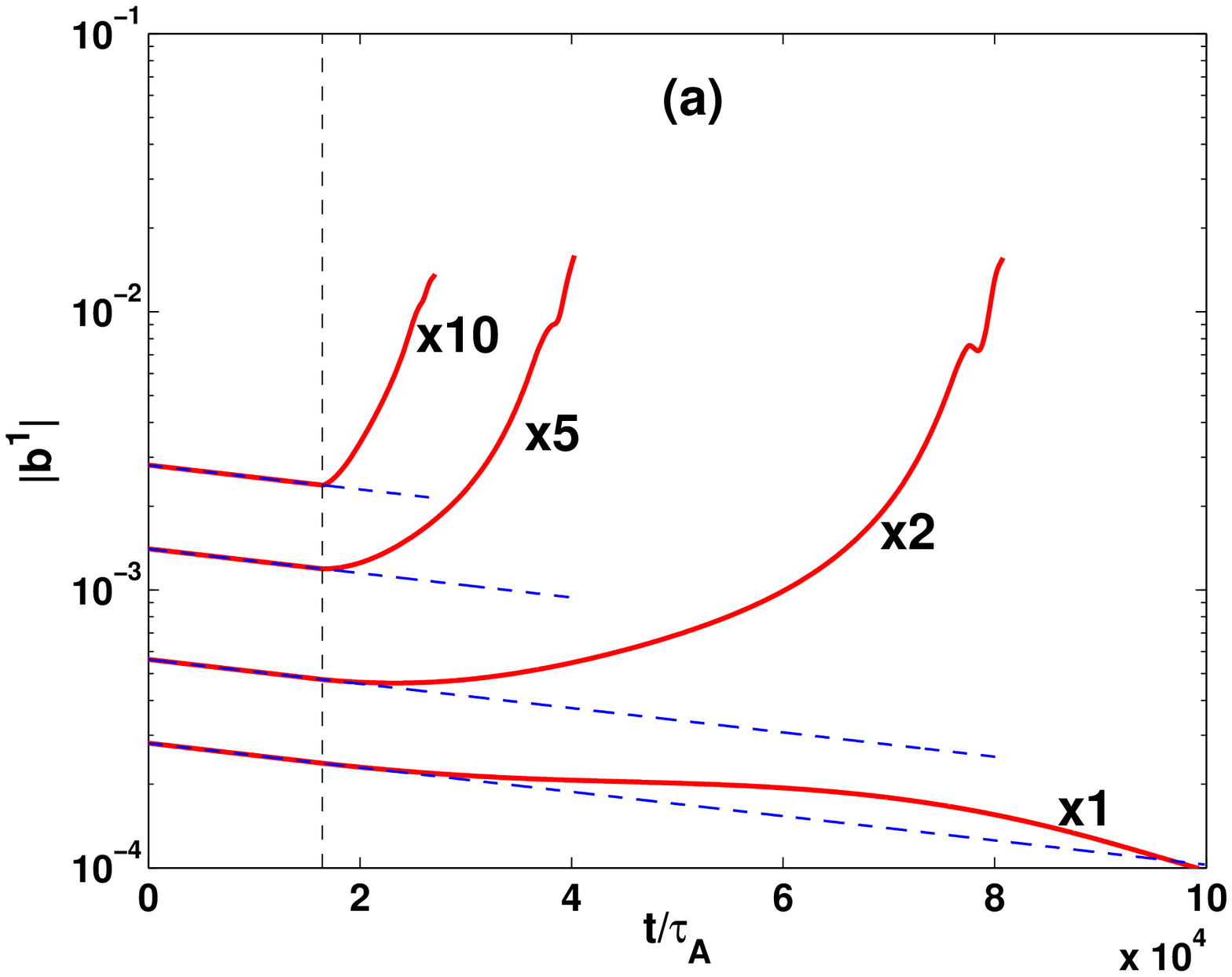}
\includegraphics[width=6.5cm]{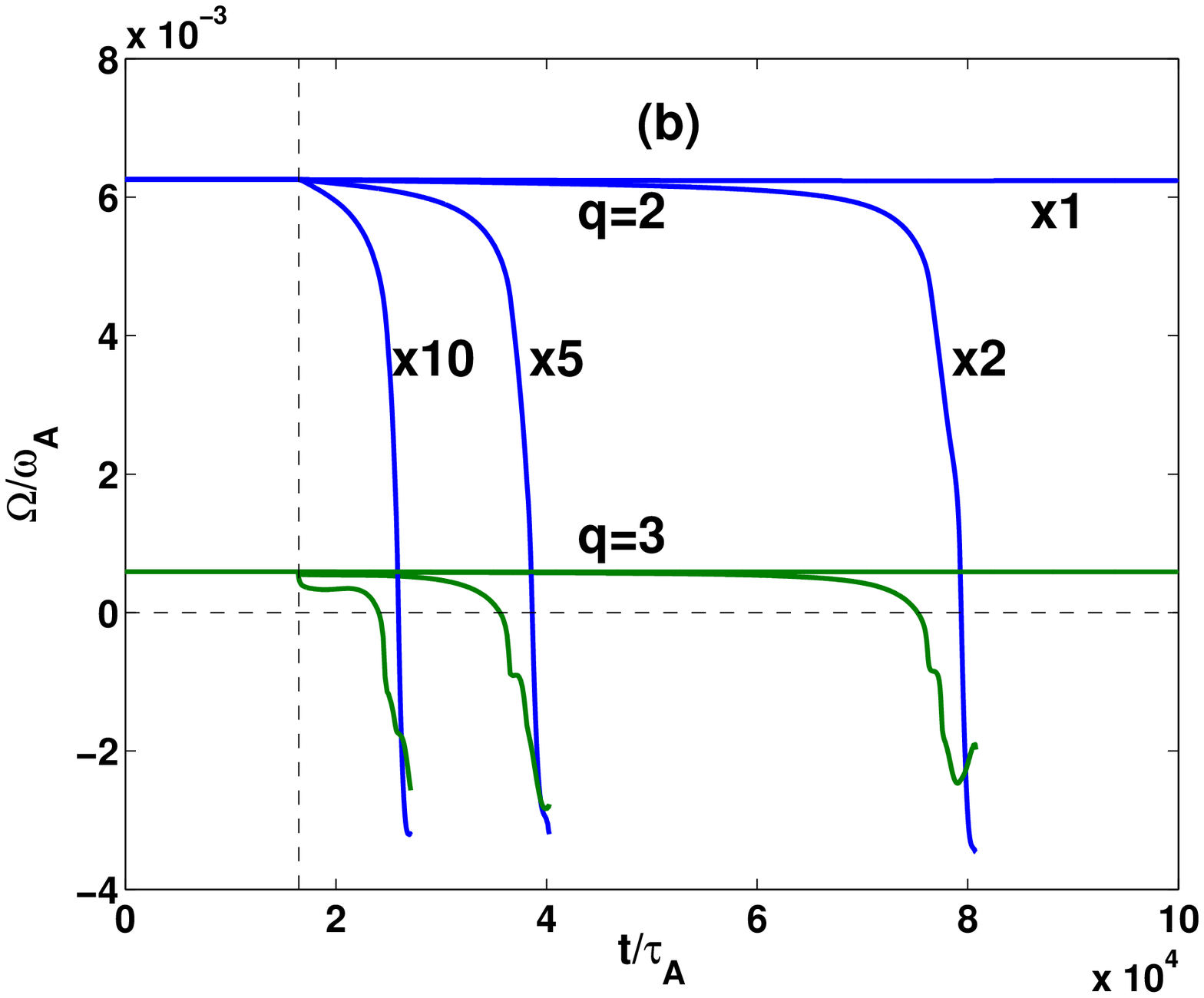}
\caption{Time traces of (a) the amplitude of the perturbed radial
field $b^1$ at the $q=2$ surface, and (b) the plasma rotation frequency at
the $q=2$ and $q=3$ surfaces, with various choices of the initial mode
amplitude. The dashed curves in (a) correspond to the 
exponential decay of the initially stable linear mode. The dashed
vertical lines indicate the moment of time when the non-linear
coupling between the mode and the plasma flow is switched on. Both the
electromagnetic and the NTV torques are 
included in the simulations.}  
\label{fig:qlsb1all}
\end{center}
\end{figure}

As a final comparison, we plot time traces of the mode amplitude and
the rotation frequencies at rational surfaces, in
Figs. \ref{fig:qlsb1all}(a) and (b), respectively. The initial
mode amplitude is increased by factors of 2,5,10, starting from the
lowest level of $|b^1|(q=2,t=0)=2.8\times10^{-4}$. Both electromagnetic
and NTV torques are included here. We observe a progressively earlier
time of the rotational braking and the mode onset, as the initial mode
amplitude is increased. The critical mode amplitude, above which the
non-linearly coupled system bifurcates from the stable solution to an
unstable solution, is between $1.4-2.8\times10^{-4}$ for $|b^1|(q=2)$
at $t=0$, for the toroidal plasma considered in this work.    

\section{Conclusion and discussion}
We have presented toroidal simulation results of the non-linear
interaction between the RWM and the toroidal plasma flow, for a
typical advanced tokamak plasma. By tuning the amplitude of the
initial plasma flow, the linear RWM, from which we launch the
non-linear simulation, can be either stable or unstable. An initially
unstable RWM brakes the plasma flow, leading to a non-linearly more
unstable mode, compared to the linear phase. An initially stable RWM
can also brake the flow, due to the resonant amplification
effect. Depending on the initial amplitude of the mode perturbation,
the non-linear coupling can result in an either stable or unstable
solution. The critical (stable) mode amplitude, above which an
eventual rotation braking and an unstable mode onset occur, is
between $1.4-2.8\times10^{-4}$ (normalized by the toroidal vacuum
field) for the radial field at the $q=2$ surface, for our toroidal example.  

The general features of these toroidal simulation results have been
qualitatively predicted by cylindrical theory
\cite{FitzpatrickPP02,GimblettPP04}. On the other 
hand, toroidal simulations provide more quantitative answers, in
particular with regard to the bifurcation amplitude of an initially stable
RWM. Also, the change of the radial profile of the flow, which in turn
modifies the RWM eigenstructure, is normally not captured by analytic
theory. Another interesting observation, which has not been addressed in
previous theory, is the dominant role played by the NTV torque in the
flow damping due to the RWM. The net toroidal NTV torque, acting on
the plasma due to the presence of the mode as well as a resistive wall, is
generally larger than the electromagnetic ${\bf j}\times{\bf b}$
torque, independent of whether the RWM is initially stable or
unstable.

In this work, an initially stable RWM is achieved by a strong parallel
sound wave damping, in combination with the Alfv\'en and sound wave
continuum damping, at a sufficiently fast plasma flow. The parallel
sound wave damping model is often a crude approximation of the ion Landau
damping, but nevertheless helps to suppress the mode, providing
interesting toroidal cases for studying the non-linear interaction between the
RWM and the plasma flow.

In the future, it is certainly desirable to investigate the RWM-flow
coupling with more physics based mode damping
models. One such model is the drift kinetic damping included into the
MARS-K code \cite{LiuPP08}. Because of the often rather complicated
dependence of the linear mode stability on the plasma flow speed within
the kinetic model, we expect generally more rich phenomena in the
non-linear coupling between the mode and the plasma rotation
braking. In particular, at slow plasma flow, the kinetic effects may introduce
significant changes in the time evolution of the rotation
profile. This is because, at sub-diamagnetic flow,  the precessional
drift resonances of trapped 
particles can become important for the RWM stability. In a
self-consistent model, the drift kinetic effect can also 
modify the eigenfunction of the fluid RWM. This adds even more interesting
aspects into the non-linear interaction between the mode and the
plasma flow. 
             
As far as the RWM is concerned, there are other interesting aspects that
should be addressed in the future work, such as the effects of 3D
conducting structures, the presence of electromagnetically thick walls (e.g. the
volumetric blanket modules in ITER), the presence of field errors that
have been included in cylindrical theory
\cite{GimblettPP04,FitzpatrickPP07}. Finally, quantitative 
simulations of the RWM interaction with the plasma flow for ITER
plasmas can be performed, based on a similar approach as that reported in this work.

{\bf Acknowledgments.}
This work was part-funded by the RCUK Energy Programme under grant
EP/I501045 and the European Communities under the contract of
Association between EURATOM and CCFE. The views and opinions expressed
herein do not necessarily reflect those of the European Commission.

Youwen Sun would like to acknowledge the support from the National
Magnetic Confinement Fusion Science Program of China under Grant
No. 2013GB102000 and No. 2012GB105000, and the National Natural
Science Foundation of China under Grant No. 11205199 and No. 10725523.  

%\section*{References}

\newpage

\begin{figure}
\begin{center}
\includegraphics[width=10cm]{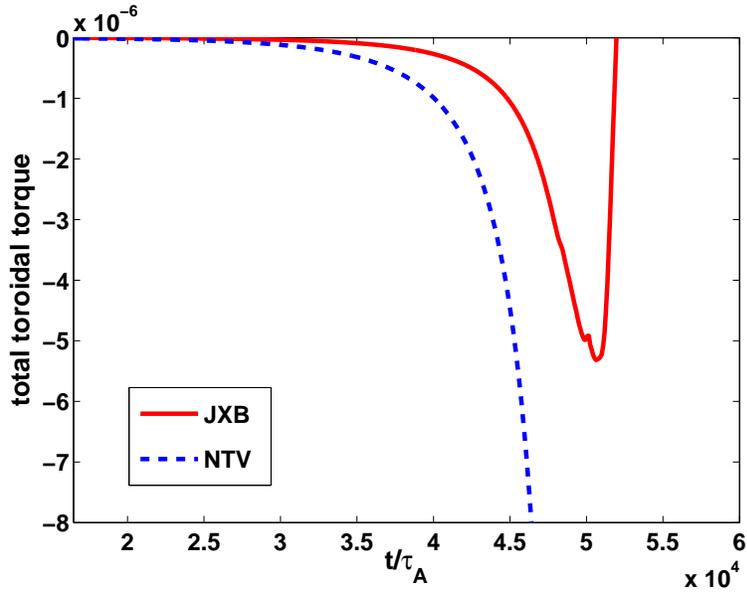}
\caption{Time trace of the net toroidal electromagnetic (solid line)
and NTV (dashed line) torques acting on the plasma, during the
non-linear evolution of an initially 
unstable RWM as described in Fig. \ref{fig:qlu}.}  
\label{fig:qlutorq}
\end{center}
\end{figure}

\end{document}